\documentclass[11pt]{article}
\usepackage[utf8]{inputenc}
\usepackage[T1]{fontenc}
\usepackage{amssymb,amsfonts,amsmath,amsthm}
\usepackage{latexsym}
\usepackage[english]{babel}
\usepackage[colorlinks=true]{hyperref}
\usepackage{mathrsfs}
\usepackage{multirow}
\usepackage{hhline}
\usepackage{longtable}
\usepackage{cite}
\usepackage{authblk}

\newtheorem{theorem}{Theorem}

\newtheoremstyle{Definition}
	{3pt}			%	space above
	{3pt}			%	space below
	{}				%	body font
	{\parindent}	%	indent emount	
	{\scshape}		%	Theorem head font
	{.}				%	Punctuation after theorem head
	{.5em}			% 	Space after theorem head
	{}				%	Theorem head spec (can be left empty, meaning `normal')
\theoremstyle{Definition}
\newtheorem{remark}{Remark}
\newtheorem{definition}{Definition}

\newcommand{\A}{\mathscr{A}}
\newcommand{\p}{\partial}
\newcommand{\g}{\mathscr{G}}
\newcommand{\F}{\mathscr{F}}
\newcommand{\dS}{\mathrm{dS}_3}
\DeclareMathOperator{\rank}{\mathrm{rank}}
\DeclareMathOperator{\ind}{\mathrm{ind}}
\DeclareMathOperator{\Exp}{\mathrm{Exp}}

\title{\bf Exact solutions of the Klein-Gordon equation in external electromagnetic fields on 3D de Sitter background}

\author[1]{Alexey A. Magazev\thanks{magazev@omgtu.ru}}
\author[1]{Maria N. Boldyreva\thanks{b\_oldyrev\_a@mail.ru}}
\affil[1]{Omsk State Technical University, Omsk, Russia}
\date{}

\begin{document}
\maketitle

\abstract{
In this study, we investigate the symmetry properties and the possibility of exact integration of the Klein--Gordon equation in the presence of an external electromagnetic field on 3D de Sitter background. 
We present an algorithm for constructing the first-order symmetry algebra and describe its structure in terms of Lie algebra extensions. 
Based on the well-known classification of the inequivalent subalgebras of the algebra $\mathfrak{so}(1,3)$, we obtain the classification of the electromagnetic fields on $\dS$ admitting first-order symmetry algebras of the Klein--Gordon equation. 
Then, we select the integrable cases, and for each of them, we construct exact solutions, using the non-commutative integration method developed by Shapovalov and Shirokov.
In Appendix, we present an original algebraic method for constructing the special local coordinates on de Sitter space, in which the basis vector fields for subalgebras of the algebra $\mathfrak{so}(1,3)$ have the simplest form.}

\section*{Introduction}

One of the most important problems of modern theoretical and mathematical physics is the problem of finding exact solutions of relativistic wave equations in external fields.
In quantum electrodynamics, for example, exact solutions of the Dirac and Klein--Gordon equations allow us to construct the so-called Furry picture. In the framework of this picture, the interaction with an external field is described exactly, and the interaction with a quantized photon field is  taken into account using perturbation theory~\cite{Fur51}. 
Exact solutions of the Dirac and Klein--Gordon equations are also extremely useful in studying various vacuum quantum effects in strong external fields, where the methods and approaches of the standard perturbation theory are not applicable (see, for instance, \cite{BirDev84, GriMamMos94}).

This study focuses on exact solutions of the Klein--Gordon equation in the presence of an external electromagnetic field on 3D de Sitter background $\dS$. When writing this paper, we mainly followed two objectives. First of all, we would like to demonstrate the potential of a new powerful method for constructing exact solutions of relativistic wave equations called \textit{the method of noncommutative integration}. This method was developed in the early 90s of the last century by Shapovalov and Shirokov~\cite{ShaShi95}.
In contrast to the well-known method of separation of variables, the application of which is associated with the study of commutative subalgebras of the first- and second-order symmetry operators~(see, for instance, \cite{Mil77, Woo75, Sha79, Ben16}), the method of noncommutative integration allows us to construct exact solutions using only the algebra of the first-order symmetry operators, which is noncommutative in general~\cite{BagBalGitShi02, Kli01, Mag12}.
Secondly, in the paper, we would like to give the exhaustive classification of electromagnetic fields on $\dS$ that admit the existence of a first-order symmetry algebra for the Klein-Gordon equation and, using this algebra, to construct the corresponding exact solutions in cases where this is possible.

Note that $n$-dimensional de Sitter space $\mathrm{dS}_n$ is regularly in the center of the attention of specialists in mathematical and theoretical physics, which is mainly owing to two reasons.
On the one side, from the modern viewpoint, it is considered that, in the four-dimensional case, this space describes the very early stages of the expansion of the universe quite accurately.
On the other side, the space $\mathrm{dS}_n$ is a maximally symmetric Lorentzian manifold satisfying Einstein's equations with a positive cosmological constant.
In particular, the high symmetry of this space allows one to integrate relativistic wave equations and, as a consequence, to investigate in detail of various quantum effects on its background~\cite{BirDev84, Ort04, All85, Yag09}. 
The presence of an external electromagnetic field, however, can cause the symmetry breaking that makes the problem of classification of electromagnetic fields admitting exact solutions of relativistic wave equations relevant.
We note that although the problem of integrating free relativistic equations on de Sitter background has long been solved (see, for instanse, \cite{BirDev84, Otc85, Pol89}), exact solutions of these equations in external electromagnetic fields are still poorly systematized.
At the same time, such exact solutions are extremely important, for instance, in connection with the study of the particle creation effect in an external electromagnetic field on a curved spacetime~\cite{Gar94, Vil95, Mor09, BavKimSta18}.

The structure of this paper is as follows. In Section \ref{sec:1}, we recall the necessary information about the algebra of the first-order symmetry operators for the Klein--Gordon equation. 
Based on the defining equations for these operators (see \cite{Car77, Van07}), we present an algorithm for constructing this algebra and describe its structure in terms of Lie algebra extensions.
In Section \ref{sec:2}, we outline some geometric and group properties of 3D de Sitter space as well as the well-known classification of all inequivalent subalgebras of the algebra $\mathfrak{so}(1,3)$, which is the Lie algebra of the isometry group of $\dS$.
Using this classification, in Section \ref{sec:3}, we obtain the list of \textit{all} electromagnetic fields on $\dS$ that admit nontrivial first-order symmetry algebras of the Klein--Gordon equation. In the same section, we explicitly construct these algebras and describe their structures by writing down the corresponding commutation relations.
In Section \ref{sec:4}, we examine the integrability problem for the Klein--Gordon equation in external electromagnetic fields on $\dS$. 
We emphasize that by \textit{integrability} we mean that the Klein-Gordon equation can be constructively solved by the process of its reduction to an ordinary differential or algebraic equation (as it is understood in the theory of separation of variables).
Using the condition of noncommutative integrability \cite{ShaShi95}, we find all the integrable cases, and, for each of them, we reduce the original Klein--Gordon equation to an auxiliary ordinary differential equation. In addition, we express the general solution of the auxiliary equation in terms of known special functions where possible.

All local constructions are presented in special local coordinates on $\dS$, in which the basis vector fields for subalgebras of $\mathfrak{so}(1,3)$ have the simplest form.
In Appendix, the original algebraic method for constructing such coordinates is described.

\section{First-order Klein--Gordon symmetry operators}
\label{sec:1}

Let $(M,g)$ be a smooth $n$-dimensional Lorentzian manifold of signature $(+,-,\dots, -)$, $x^1, \dots, x^n$ are local coordinates  on $M$.
We consider the covariant Klein--Gordon equation for a charged massive scalar field in the presence of an external electromagnetic field~\cite{BirDev84, GriMamMos94}:
\begin{equation}
\label{KGeq}
\hat{H} \varphi \equiv \left ( g^{ab} D_a D_b + \zeta R + m^2 \right ) \varphi = 0.
\end{equation}
Here, $\varphi$ denotes the scalar field, $g^{ab}$ are the contravariant components of the metric $g$, $R$ is the Riemannian scalar curvature, $m$ and $e$ are the mass and charge of the field quanta, respectively.
The generalized covariant derivatives $D_a$ are defined as $D_a \equiv \nabla_a - i e \A_a$, where 
$\nabla_a$ is the covariant derivative with respect to the coordinate vector field $\p_{x^a} \equiv \p/ \p x^a$, and $\A_a$ are the components of the electromagnetic potential 1-form $\A$.
The dimensionless parameter $\zeta$ can take two values: $\zeta = 0$ (the minimally coupled case) and $\zeta = ( n - 1 ) / ( 4 n )$ (the conformally coupled case).
In what follows we will use the Einstein summation convention unless stated otherwise. 
Latin indices are raised and lowered by the metric $g$.

An operator $\hat{X}$ is said to be a \textit{symmetry operator} for Eq.~\eqref{KGeq} if the following condition is satisfied
\begin{equation}
\label{comm}
[ \hat{H}, \hat{X}] \equiv \hat{H} \hat{X} - \hat{X} \hat{H} = \hat{Q} \hat{H},
\end{equation}
where $\hat{Q}$ is an operator which depends on $\hat{X}$ in general.
It is known that the symmetry operators map solutions of Eq.~\eqref{KGeq} to other solutions, and the set of all symmetry operators $\mathfrak{M}$ forms a Lie algebra called the \textit{symmetry algebra} of the Klein-Gordon equation~\cite{Mil77}.

The main object of our investigation is a subalgebra of $\mathfrak{M}$ generated by all the  differential operators of the form
\begin{equation}
\label{opX}
\hat{X} = X^a(x) D_a + i e \chi(x).
\end{equation}
Here, $X^a(x)$ and $\chi(x)$ are assumed to be smooth real functions on $M$, possibly 
not everywhere defined.
Obviously, the set of all such operators forms a certain subalgebra in $\mathfrak{M}$, which we denote as $\hat{\g}$.

The conditions under which the operator \eqref{opX} is a symmetry operator of the Klein--Gordon equation are well known~(see, for instance, \cite{Car77}).
They can be obtained by substituting \eqref{opX} into \eqref{comm} and equating coefficients of 
$D_a D_b$, $D_a$, and $1$ on both sides of the resulting relation. 
As a result, we obtain the system of equations
\begin{equation}
\label{DefEq1}
\nabla^a X^b + \nabla^b X^a = 0,
\end{equation}
\begin{equation}
\label{DefEq2}
X^b \F_{ab} - \nabla_a \chi = 0,
\end{equation}
where $\F_{ab}$ are the components of the electromagnetic tensor defined by
\begin{equation*}
\F_{ab} = \frac{\p \A_b}{\p x^a} - \frac{\p \A_a}{\p x^b}.
\end{equation*}
Also, note that, for a massive scalar field, there should be $\hat{Q} = 0$.

The system of Eqs. \eqref{DefEq1} and \eqref{DefEq2} is a necessary and sufficient condition for $\hat{X}$ to be a symmetry operator for the Klein--Gordon equation.
In particular, Eq.~\eqref{DefEq1} means that $X = X^a \p_{x^a}$ is a \textit{Killing vector} of the Lorentzian manifold $(M,g)$. 
Killing vectors reflect the geometric symmetry of the spacetime and are the infinitesimal generators of its (local) isometry group $\mathrm{Iso}(M,g)$.
The set of all Killing vectors of $(M,g)$ forms a Lie algebra (with respect to the commutator of vector fields), which we denote as $\mathfrak{iso}(M,g)$.

We note that not every Killing vector $X = X^a \p_{x^a}$ from $\mathfrak{iso}(M,g)$ can be extended to a symmetry operator $\hat{X} = X^a D_a + i e \chi$ of Eq.~\eqref{KGeq}. 
Carter showed \cite{Car77} that a scalar function $\chi$ of the required form locally exists if and only if
\begin{equation}
\label{LieF}
\mathscr{L}_X \F = 0,
\end{equation}
where $\mathscr{L}_X$ denotes the Lie derivative along the vector field $X$.
Indeed, if we rewrite Eq.~\eqref{DefEq2} in terms of differential forms
\begin{equation}
\label{EqChi}
d \chi = - i_X \F,
\end{equation}
where $i_X \F \equiv \F_{ab} X^a d x^b$ is the interior product of the 2-form $\F = \frac{1}{2}\, \F_{ab}\, d x^a \wedge dx^b$ and the vector field $X = X^a \p_{x^a}$, then, it is easy to see that the local existence of the function $\chi$ is equivalent to the statement that the 1-form $-i_X \F$ is closed. 
This, in turn, is equivalent to Eq.~\eqref{LieF}, in view of the Cartan formula $\mathscr{L}_X = i_X d + d i_X$ and the condition $d\F = 0$.

Since the correspondence $X \to \mathscr{L}_X$ is a homomorphism of Lie algebras, the set of all Killing vectors satisfying the condition \eqref{LieF} for a given electromagnetic field forms a subalgebra $\g$ of the Lie algebra $\mathfrak{iso}(M,g)$.
We will call $\g$ the \textit{admissible subalgebra}.

Let $X_A = X_A^a(x) \p_{x^a}$ be Killing vectors forming a basis of~$\g$, $A = 1, \dots, \dim \g$.
Since $\g$ is a subalgebra, we have
\begin{equation}
\label{commX}
[X_A, X_B] = C_{AB}^C X_C,\quad
A, B = 1, \dots, \dim \g.
\end{equation}
Here, the constants $C_{AB}^C$ are the \textit{structure constants} of~$\g$.
By the definition of the admissible subalgebra, for each vector field $X_A$, there exists a local function $\chi_A$ satisfying the condition~\eqref{EqChi}.  
Note that this function is not uniquely defined; instead of $\chi_A$, one can choose the function $\chi_A' = \chi_A + \lambda_A$, where $\lambda_A$ is an arbitrary constant. 
The meaning of this ambiguity will be explained below.

By construction, the first-order differential operators
\begin{equation}
\label{opX_A}
\hat{X}_A = X_A^a(x) D_a + i e \chi_A,\quad
A = 1, \dots, \dim \g,
\end{equation}
commute with the operator $\hat{H}$, i.e., $\hat{X}_A$ are symmetry operators for the Klein--Gordon equation~\eqref{KGeq}.
In the general case, however, the linear span of these operators does not form the whole algebra~$\hat{\g}$. 
Indeed, using the commutation relations $[D_a, D_b] = - i e \F_{ab}$ and \eqref{commX}, the commutator of $\hat{X}_A$ and $\hat{X}_B$ can be written as \cite{Mag12}
\begin{equation}
\label{comm_opX}
[\hat{X}_A, \hat{X}_B] = C_{AB}^C \hat{X}_C + \mathbf{F}_{AB} \hat{X}_0,
\end{equation}
where
\begin{equation}
\label{FF}
\mathbf{F}_{AB} \equiv \F(X_A, X_B) - C_{AB}^C \chi_C.
\end{equation}
Here, we have introduced the trivial symmetry operator $\hat{X}_0 \equiv i e$.
In Ref.~\cite{Mag12}, it is shown that the quantities~$\mathbf{F}_{AB}$ are constant on $M$ and satisfy the identity
\begin{equation*}
\label{CocycleCond}
C_{AB}^D \mathbf{F}_{CD} + C_{BC}^D \mathbf{F}_{AD} + C_{CA}^D \mathbf{F}_{BD} = 0.
\end{equation*}
This allows us to interpret the quantities \eqref{FF} as components of a \textit{cocycle} of the admissible subalgebra $\g$ with values in the trivial $\g$-module $\mathbb{R}$ (for a review of basic results in Lie algebra cohomology and their applications in mathematical physics see \cite{de1998lie, de1998introduction}). 
Moreover, from the commutation relations \eqref{comm_opX}, it follows that the algebra $\hat{\g}$ is a one-dimensional central extension of the Lie algebra $\g$ corresponding to the cocycle~\eqref{FF}. 

As we have already noted, the functions $\chi_A$ are determined by Eq.~\eqref{EqChi} up to the addition of arbitrary constants.
As can be seen from \eqref{opX_A}, the transformation $\chi_A \to \chi_A' = \chi_A + \lambda_A$ implies the transformation $\hat{X}_A \to \hat{X}'_A = \hat{X}_A + \lambda_A \hat{X}_0$, which is a change of basis in the algebra $\hat{\g}$.
In the general case, it leads to a change of the commutation relations \eqref{comm_opX}, because the quantities \eqref{FF} are transformed as follows:
\begin{equation}
\label{Fchange}
\mathbf{F}_{AB} \to \mathbf{F}'_{AB} = \mathbf{F}_{AB} - C_{AB}^C \lambda_C.
\end{equation}

\begin{remark}
In some cases, it may be possible to choose the constants $\lambda_C$ in Eq.~\eqref{Fchange} such that all quantities $\mathbf{F}'_{AB}$ vanish.
Cocycles of this kind are called \textit{trivial cocycles} or \textit{coboundaries}.
For example, any cocycle of a semisimple Lie algebra is trivial; this result is the content of Whitehead's second lemma~\cite{Jac79}.
In the case of a trivial cocycle, the structure of the algebra $\hat{\g}$ is especially simple, since in this situation the symmetry operators $\hat{X}_A$ can be chosen so that
\begin{equation*}
[\hat{X}_A, \hat{X}_B] = C_{AB}^C \hat{X}_C.
\end{equation*}
This means that $\hat{\g}$ is isomorphic to the direct sum of the algebras $\g$ and $\mathbb{R}$: $\hat{\g} \simeq \g \oplus \mathbb{R}$.
\end{remark}

The above results allow us to list all electromagnetic fields admitting first-order symmetry operators of Eq.~\eqref{KGeq} for a given Lorentzian manifold $(M,g)$.
It is preliminary convenient to introduce the following equivalence relation.
Two electromagnetic fields with closed 2-forms $\F$ and $\F'$ are called \textit{equivalent} if they are connected by an isometry:
\begin{equation*}
\label{EquvF}
\F = \tau^* \F',\quad
\tau \in \mathrm{Iso}(M,g).
\end{equation*}
It is easy to show that the admissible subalgebras $\g$ and $\g'$ corresponding to the equivalent electro\-magnetic fields $\F$ and $\F'$ are conjugate in $\mathfrak{iso}(M,g)$ by the transformation $\tau$: $\g' = \tau \g \tau^{-1}$.
Conversely, if the subalgebras $\g$ and $\g'$ are $\tau$-conjugate, then the spaces of closed 2-forms on $M$ invariant under the subalgebras $\g$ and $\g'$ are connected by the transformation $\tau^*$.
Thus, in order to describe fully the electromagnetic fields that admit a nontrivial first-order symmetry operator of Eq.~\eqref{KGeq}, one may use the list of subalgebras of the algebra $\mathfrak{iso}(M,g)$ up to conjugations.
For each subalgebra $\g = \{ X_A \}$ from this list, we can find the most general form of a closed 2-form $\F$ such that\footnote{Note that the classes of electromagnetic fields corresponding non-conjugate subalgebras of $\mathfrak{iso}(M,g)$ may have nonzero intersections.}
\begin{equation}
\label{LieFA}
\mathscr{L}_{X_A} \F = 0,\quad
A = 1, \dots, \dim \g,
\end{equation}
and then explicitly construct a basis of the Lie algebra $\hat{\g}$ in accordance with the formula \eqref{opX_A}.

\section{The algebra of Killing vectors of $\dS$ and its inequivalent sub\-algebras}
\label{sec:2}

De Sitter space $\dS$ is the three-dimensional one-sheeted hyperboloid in four-dimensional Minkowski space $\mathbb{R}^{1,3}$, described by the equation~\cite{BirDev84}
\begin{equation}
\label{deSit}
x_0^2 - x_1^2 - x_2^2 - x_3^2 = - \alpha^2.
\end{equation} 
Here, $\alpha$ is a nonzero positive constant with units of length called the \textit{de Sitter radius}.
The metric of $\dS$ induced by the standard Minkowski metric $\eta_{ij} = \mathrm{diag}(1,-1,-1,-1)$ is non-degenerate and has the Lorentzian signature $(+,-,-)$.
From the group-theoretic point of view, three-dimensional de Sitter space is the homogeneous space $O(1,3)/O(1,2)$, where $O(1,n)$ denotes the pseudo-orthogonal group of all linear transformations that leave invariant a non-degenerate quadratic form of signature $(1,n)$.
From the topological point of view, $\dS$ is the direct product of the real line and the two-dimensional sphere: $\dS = \mathbb{R} \times S^2$.
From now on, we assume that $\alpha = 1$.
The scalar curvature of $\dS$, in this case, is equal to $R = 6$.

The isometry group of de Sitter space $\dS$ is the Lorentz group $O(1,3)$.
The identity component $SO^+(1,3)$ of this group, called the \textit{restricted Lorentz group}, is generated by the six Killing vectors: 
\begin{equation}
\label{so13basis}
J_{ij} = x_i\, \frac{\p}{\p x^j} - x_j \frac{\p}{\p x^i},\quad
0 \leq i < j \leq 3.
\end{equation}
(Raising and lowering the indices are performed by the metric $\eta_{ij}$).
The Lie algebra $\mathfrak{so}(1,3)$ defined by these vector fields has the following commutation relations:
\begin{equation*}
\label{so13comm}
[J_{ij}, J_{kl}] = \eta_{jk} J_{il} - \eta_{ik} J_{jl} + \eta_{il} J_{jk} - \eta_{jl} J_{ik},\quad
i,j = 0, 1, 2, 3.
\end{equation*}
We note that the vector fields $J_{12}$, $J_{13}$, and $J_{23}$ correspond to ordinary spatial rotations, while the Killing vectors $J_{01}, J_{02}$, and $J_{03}$ are associated with the Lorentz boosts.  

As noted in the previous section, the first step in the classification of electromagnetic fields that admit first-order symmetry operators of Eq.~\eqref{KGeq} on $\dS$ is to list all inequivalent subalgebras of the algebra~$\mathfrak{so}(1,3)$. 
Every such subalgebra $\g \subset \mathfrak{so}(1,3)$ is fixed by a set of $n$ linearly independent vector fields $ X_A = \sum_{i<j} C_A^{ij} J_{ij}$, which are some linear combinations of the Killing vectors~\eqref{so13basis}.

All inequivalent proper subalgebras of the algebra $\mathfrak{so}(1,3)$ are well known~(see, for instance, \cite{FriWin64, BarBarFus91, Hal04}); their exhaustive list (up to conjugation) is given in Table~\ref{tab:1}. 
In this table, we label each inequivalent subalgebra $\g_{n,m}$ by its dimension $n$ and its number $m$ in the list.
If there is a one-parametric family of subalgebras, we denote this family as $\g^a_{n,m}$, where $a$ is a parameter.
The third column of the table shows the nonzero commutation relations between the basis vectors of the subalgebras.

\begin{table}[!h]
\caption{\small{Inequivalent subalgebras of~$\mathfrak{so}(1,3)$.}}
\label{tab:1}
\small
\begin{center}
\begin{tabular}{|c|c|c|}
\hhline{===}
Subalgebra	&	Infinitesimal generators $X_A$	& Commutation relations	\\
\hhline{===}
$\g_{1,1}$	&	$X_1 = J_{03}$	&	---	\\
\hline
$\g_{1,2}$	&	$X_1 = J_{12}$	&	---	\\
\hline
$\g_{1,3}^a$&	$X_1 = J_{12} + a J_{03}$, $a > 0$	&	---	\\
\hline
$\g_{1,4}$	&	$X_1 = J_{13} - J_{01}$	&	---	\\
\hline
$\g_{2,1}$	&	$X_1 = J_{13} - J_{01}$, $X_2 = J_{23} - J_{02}$	&	---	\\
\hline
$\g_{2,2}$	&	$X_1 = J_{12}$, $X_2 = J_{03}$	&	---	\\
\hline
$\g_{2,3}$	&	$X_1 = J_{13} - J_{01}$, $X_2 = J_{03}$	&	$[X_1, X_2] = - X_1$	\\
\hline
$\g_{3,1}$	&	$X_1 = J_{13} - J_{01}$, $X_2 = J_{23} - J_{02}$, $X_3 = J_{03}$	&	$[X_1, X_3] = - X_1,\ [X_2, X_3] = - X_2$	\\
\hline
$\g_{3,2}$	&	$X_1 = J_{13} - J_{01}$, $X_2 = J_{23} - J_{02}$, $X_3 = J_{12}$	&	$[X_1, X_3] = X_2,\ [X_2, X_3] = - X_1$	\\
\hline
\multirow{2}{*}{$\g_{3,3}^a$}&	$X_1 = J_{13} - J_{01}$, $X_2 = J_{23} - J_{02}$, 	&	$[X_1, X_3] = X_2 - a X_1$,	\\
			&	$X_3 = J_{12} + a J_{03}$, $a > 0$	& $[X_2, X_3] = - X_1 - a X_2$	\\
\hline
\multirow{2}{*}{$\g_{3,4}$}	&	\multirow{2}{*}{$X_1 = J_{12},\ X_2 = J_{13},\ X_3 = J_{23}$}	&	$[X_1, X_2] = X_3,\ [X_1, X_3] = - X_2$,	\\
			&	&	$[X_2, X_3] = X_1$	\\
\hline
\multirow{2}{*}{$\g_{3,5}$}	&	\multirow{2}{*}{$X_1 = J_{01},\ X_2 = J_{02},\ X_3 = J_{12}$}	&	$[X_1, X_2] = X_3,\ [X_1, X_3] = X_2$,	\\
			&	&	$[X_2, X_3] = - X_1$\\
\hline			
\multirow{2}{*}{$\g_{4,1}$}	&	$X_1 = J_{13} - J_{01}$, $X_2 = J_{23} - J_{02},\ X_3 = J_{12}$,	&	$[X_1, X_3] = X_2$, $[X_1, X_4] = - X_1$,\\
			& $X_4 = J_{03}$	& $[X_2, X_3] = - X_1$, $[X_2, X_4] = - X_2$	\\
\hline
\end{tabular}
\end{center}
\end{table}

Let us give some explanations for the subalgebras listed in Table \ref{tab:1}.

The one-dimensional subalgebras $\g_{1,1}$, $\g_{1,2}$, and $\g_{1,4}$ generate one-parameter subgroups of Lorentz boosts $SO(1,1)$, rotations $SO(2)$, and null rotations $SO(0,1)$, respectively.
The family of one-dimen\-sional subalgebras $\g^a_{1,3}$ parametrized by the positive parameter $a$ generates the family of one-parameter loxodromic transformations.

The subalgebra $\g_{2,1}$ is associated with a two-dimensional Abelian group of transformations consisting of zero rotations.
The subalgebra $\g_{2,2}$ corresponds to a two-dimensional Abelian transformation group consisting of Lorentz boosts (in the $x^3$ direction), rotations in the plane $x^1 x^2$, and their combinations.
The subalgebra $\g_{2,3}$ generates a non-Abelian two-dimensional group isomorphic to
 the affine group~$A(1)$.

The subalgebra $\g_{3,1}$ is the Lie algebra of Bianchi type V; the corresponding group is isomorphic to the group of Euclidean homotheties~$\mathrm{Hom}(2)$.
The subalgebra $\g_{3,2}$ is the Lie algebra of Bianchi type VII$_0$; it is isomorphic to the Lie algebra of the euclidean group~$E(2)$.
The subalgebra $\g_{3,3}^a$, where $a>0$, is of Bianchi type VII$_a$. 
The subalgebra $\g_{3,4}$ is of Bianchi type IX; the corresponding transformation group is the rotation group $SO(3)$.
Finally, the subalgebra $\g_{3,5}$ of Bianchi type VIII is isomorphic to the Lie algebra of the group $SL(2, \mathbb{R})$.

There is only one (up to conjugation) four-dimensional subalgebra $\g_{4,1}$; it is isomorphic to the Lie algebra of the group of Euclidean similitudes $\mathrm{Sim}(2)$.

There are no five-dimensional subalgebras of the Lie algebra $\mathfrak{so}(1,3)$.

\section{Electromagnetic fields admitting nontrivial first-order symmetry algebras on $\dS$ }
\label{sec:3}

For subsequent purposes, we need to fix a local coordinate system in the space $\dS$ and rewrite the basis vector fields $X_A$ for each of the subalgebras $\g_{n,m} \subset \mathfrak{so}(1,3)$ in these coordinates. 
Since our problem is to find a closed 2-form $\F$ satisfying Eq.~\eqref{LieFA}, it is reasonable to use a coordinate system in which  the vector fields $X_A$ have the simplest form.
In this sense, local coordinates that ``rectify'' the integral submanifolds of the system of vector fields $X_1, \dots, X_n$ are most suitable. 
Let us recall the rigorous definition of such coordinates in the context of the situation under consideration.

Let $\g_{n,m}$ be a subalgebra from Table~\ref{tab:1}, $X_1, \dots, X_n $ are its basis vector fields. 
Denote by $r$ the dimension of the space spanned by the vectors $X_1|_x, \dots, X_n|_x$ for a point $x \in \dS$ in general position, i. e.
\begin{equation}
\label{rdef}
r \equiv \sup \limits_{x \in \dS} \mathrm{rank}\, \| X_A^a(x) \|.
\end{equation}
If $x_0 \in \dS$ is a point in general position, then, according to the well-known Frobenius theorem, there are local coordinates $(q,u) = (q^1, \dots, q^r$, $u^1, \dots, u^{3 - r})$ near of $x_0 \in \dS$ such that the integral submanifolds of $\{ X_1, \dots, X_n \}$ intersect this coordinate chart along the ``slices'' $u^1 = c_1, \dots, u^{3-r} = c_{3-r}$, where $c_1, \dots, c_{3-r}$ are arbitrary constants~(see Ref.~\cite{Olv00}).
In this case, $q^1, \dots, q^r$ are regarded as local coordinates on the ``slices'', and the vector fields $X_A$ being tangent to the ``slices'' are written as
\begin{equation*}
\label{adaptX_A}
X_A = \sum \limits_{a = 1}^r X_A^a(q, u) \frac{\p}{\p q^a},\quad
A = 1, \dots, n.
\end{equation*}
In turn, the coordinates $u^1,\dots,u^{3-r}$ are local invariants of the transformation group $G_{n,m}$ generated by the Lie algebra $\g_{n,m}$.
We call the coordinates $(q,u) = (q^1, \dots, q^r$, $u^1, \dots, u^{3 - r})$ the \textit{rectifying coordinates} associated with the subalegbras $\g_{n,m}$.

Let us now consider each of the subalgebras from Table~\ref{tab:1}.
In each case, it is not hard to construct local rectifying coordinates associated with $\g_{n,m}$ and then to find the most general form of the closed 2-form $\F$ satisfying Eq.~\eqref{LieFA}.
The results of these calculations are summarized in Table~\ref{tab:2}, in which for each subalgebra $\g_{n,m}$ we indicate the number $r$ calculated by Eq.~\eqref{rdef}, the vector fields $X_A$, and the corresponding invariant closed 2-form $\F$ in the rectifying coordinates.
In this table, $f, f_1, f_2$ denote arbitrary smooth functions of their arguments, and $\mu, \mu_1, \mu_2$ are arbitrary constants.
The functions $x^i = x^i(q^1, \dots, q^r, u^1, \dots, u^{3-r})$ defining the rectifying local coordinates associated with the subalgebras $\g_{n,m}$ as well as a constructive algebraic method for their construction are given in Appendix A. 

\begin{table}[h!]
\caption{\small Basis vector fields $X_A \in \g_{n,m}$ and the corresponding closed invariant 2-forms $\F$ in rectifying local coordinates $(q,u)$.}
\label{tab:2}
\small
\begin{center}
\begin{tabular}{|c|c|c|c|}
\hhline{====}
Subalgebra	& $r$	&	Infinitesimal generators $X_A$ 	& 	Closed 2-form $\F$, such that	\\
		&		&	 	&	$\mathscr{L}_{X_A} \F = 0$ for all $A = 1, \dots, n$	\\
\hhline{====}
$\g_{1,1}$		&	\multirow{4}{*}{1}	&	\multirow{4}{*}{$X_1 = \p_{q^1}$}	&	\multirow{4}{*}{$d q^1 \wedge d f_1(u^1, u^2) + f_2(u^1, u^2) d u^1 \wedge d u^2$}		\\
\hhline{-~~~}
$\g_{1,2}$		&	&	&	\\
\hhline{-~~~}
$\g_{1,3}^a$	&	&	&	\\
\hhline{-~~~}
$\g_{1,4}$		&	&	&	\\
\hline
$\g_{2,1}$	&	\multirow{3}{*}{2}	&	\multirow{2}{*}{$X_1 = \p_{q^1},\ X_2 = \p_{q^2}$}		&	$\mu\, d q^1 \wedge d q^2 + f_1(u^1) d q^1 \wedge d u^1 + \hfill$					\\
\hhline{-~~~}
$\g_{2,2}$	&	&	&	$ \hfill + f_2(u^1) d q^2 \wedge d u^1$	\\
\hhline{-~--}
$\g_{2,3}$	&	& $X_1 = \p_{q^1},\quad X_2 = - q^1 \p_{q^1} + \p_{q^2}$		&	$\exp(q^2) d q^1 \wedge \left ( f_1(u^1) d q^2 + d f_1(u^1) \right ) + \hfill$	\\	
			&	&	&	$\hfill + f_2(u^1) d q^2 \wedge d u^1$	\\
\hline
\multirow{2}{*}{$\g_{3,1}$}	&	\multirow{2}{*}{3}	&	$X_1 = \p_{q^1},\ X_2 = \p_{q^2},$	&	\multirow{2}{*}{$\exp(q^3) (\mu_1 d q^1 + \mu_2 d q^2) \wedge d q^3$}		\\
	&		&	$X_3 = - q^1 \p_{q^1} - q^2 \p_{q^2} + \p_{q^3}$	&	\\
\hline
$\g_{3,2}$	& 2	&	$X_1 = \p_{q^1},\ X_2 = \p_{q^2},\ X_3 = - q^2 \p_{q^1} + q^1 \p_{q^2}$	&	$\mu\, d q^1 \wedge d q^2$	\\	
\hline
\multirow{2}{*}{$\g_{3,3}^a$}	&	\multirow{2}{*}{3}	&	$X_1 = \p_{q^1},\ X_2 = \p_{q^2},$	&	$\exp(a q^3) \left ( \mu_1 \cos(q^3) + \mu_2 \sin(q^3) \right ) d q^1 \wedge d q^3 + \hfill$		\\	
	&	&	$  X_3 = -( a q^1 + q^2) \p_{q^1} + (q^1 - a q^2) \p_{q^2} + \p_{q^3} $	&	$\hfill + \exp(a q^3) \left ( \mu_1 \sin(q^3) - \mu_2 \cos(q^3) \right ) d q^2 \wedge d q^3$	\\
\hline
\multirow{3}{*}{$\g_{3,4}$}	&	\multirow{3}{*}{2}	&	$X_1 = \p_{q^1},$	&	\multirow{6}{*}{$\mu \cos(q^2) d q^1 \wedge d q^2$}\\
	&	&	$X_2 = \sin(q^1) \tan(q^2) \p_{q^1} + \cos(q^1) \p_{q^2},$	&	\\
	&	&	$X_3 = \cos(q^1) \tan(q^2) \p_{q^1} - \sin(q^1) \p_{q^2}$	&	\\
\hhline{---~}
\multirow{3}{*}{$\g_{3,5}$}	&	\multirow{3}{*}{2}	&	$X_1 = \p_{q^1},$	&	\\
	&	&	$X_2 = \sinh(q^1) \tan(q^2) \p_{q^1} + \cosh(q^1) \p_{q^2},$	&	\\
	&	&	$X_3 = \cosh(q^1) \tan(q^2) \p_{q^1} + \sinh(q^1) \p_{q^2}$	&	\\
\hhline{----}
\multirow{2}{*}{$\g_{4,1}$}	&	\multirow{2}{*}{3}	&	$X_1 = \p_{q^1},\ X_2 = \p_{q^2},\ X_3 = - q^2 \p_{q^1} + q^1 \p_{q^2}$,	&	\multirow{2}{*}{0}	\\
	&	&	$X_4 = -q^1 \p_{q^1} - q^2 \p_{q^2} + \p_{q^3}$	&	\\
\hline
\end{tabular}
\end{center}
\end{table}

By construction, the electromagnetic fields listed in Table \ref{tab:2} admit nontrivial first-order symmetry algebras of the Klein--Gordon eqiation~\eqref{KGeq}. To construct bases of these algebras explicitly, we again consider each subalgebra $\g_{n,m}$ separately.
For each vector fields $X_A \in \g_{n,m}$ from Table~\ref{tab:2}, we find the function $\chi_A$, that is, a solution of Eq.~\eqref{EqChi}, and we then write out the explicit form of the associated symmetry operator $\hat{X}_A$ according to Eq.~\eqref{opX_A}. Having constructed the basis $\{ \hat{X}_0 = i e, \hat{X}_1, \dots, \hat{X}_n \}$ of the symmetry algebra $\hat{\g}_{n,m}$, we list the nonzero commutation relations between its basis elements.

\subsection*{Symmetry algebras $\hat{\g}_{1,1}$, $\hat{\g}_{1,2}$, $\hat{\g}_{1,3}^a$ and $\hat{\g}_{1,4}$}

\begin{equation*}
\hat{X}_1 = D_1 - i e f_1(u^1,u^2).
\end{equation*}

\subsection*{Symmetry algebras $\hat{\g}_{2,1}$ and $\hat{\g}_{2,2}$}

\begin{equation*}
\hat{X}_1 = D_1 - i e \left ( \frac{1}{2}\, \mu q^2 + \int f_1(u^1) d u^1 \right ),\quad
\hat{X}_2 = D_2 + i e \left ( \frac{1}{2}\, \mu q^1 + \int f_2(u^1) d u^1 \right ),
\end{equation*}
\begin{equation*}
[\hat{X}_1, \hat{X}_2] = \mu \hat{X}_0.
\end{equation*}

\subsection*{Symmetry algebra $\hat{\g}_{2,3}$}

\begin{equation*}
\hat{X}_1 = D_1,\quad
\hat{X}_2 = - q^1 D_1 + D_2 - i e \int f_2(u^1) d u^1,
\end{equation*}
\begin{equation*}
[\hat{X}_1, \hat{X}_2] = - \hat{X}_1.
\end{equation*}

\subsection*{Symmetry algebra $\hat{\g}_{3,1}$}

\begin{multline}
\label{X31a}
\hat{X}_1 = D_1 - i e \mu_1 \exp(q^3),\quad
\hat{X}_2 = D_2 - i e \mu_2 \exp(q^3),\\
\hat{X}_3 = - q^1 D_1 - q^2 D_2 + i e \exp(q^3) \left ( \mu_1 q^1 + \mu_2 q^2 \right ),
\end{multline}
\begin{equation}
\label{commX31}
[\hat{X}_1, \hat{X}_2] = 0,\quad
[\hat{X}_1, \hat{X}_3] = - \hat{X}_1,\quad
[\hat{X}_2, \hat{X}_3] = - \hat{X}_2.
\end{equation}

\subsection*{Symmetry algebra $\hat{\g}_{3,2}$}

\begin{equation}
\label{X32}
\hat{X}_1 = D_1 - i e \mu q^2,\quad
\hat{X}_2 = D_2 + i e \mu q^1,\quad
\hat{X}_3 = - q^2 D_1 + q^1 D_2 + \frac{i}{2}\, e \mu\, \left ( (q^1)^2 + (q^2)^2 \right ),
\end{equation}
\begin{equation}
\label{commX32}
[\hat{X}_1, \hat{X}_2] = \mu \hat{X}_0,\quad
[\hat{X}_1, \hat{X}_3] = \hat{X}_2,\quad
[\hat{X}_2, \hat{X}_3] = - \hat{X}_1.
\end{equation}

\subsection*{Symmetry algebra $\hat{\g}_{3,3}^a$}

\begin{equation}
\label{X33_1}
\hat{X}_1 = D_1 + i e \exp(a q^3) \left ( \frac{\mu_2 - a \mu_1}{1 + a^2}\, \cos(q^3) - \frac{\mu_1 + a \mu_2}{1 + a^2}\, \sin(q^3) \right ),
\end{equation}
\begin{equation}
\label{X33_2}
\hat{X}_2 = D_2 + i e \exp(a q^3) \left ( \frac{\mu_2 - a \mu_1}{1 + a^2}\, \sin(q^3) + \frac{\mu_1 + a \mu_2}{1 + a^2}\, \cos(q^3) \right ),
\end{equation}
\begin{multline}
\label{X33_3}
\hat{X}_3 = - \left ( a q^1 + q^2 \right ) D_1 + \left ( q^1 - a q^2 \right ) D_2 + D_3 + 
\\
+ i e \exp(a q^3) \left ( \mu_1 q^1 - \mu_2 q^2 \right ) \cos(q^3) + i e \exp(a q^3) \left ( \mu_1 q^2 + \mu_2 q^1 \right ) \cos(q^3),
\end{multline}
\begin{equation}
\label{commX33}
[\hat{X}_1, \hat{X}_2] = 0,\quad
[\hat{X}_1, \hat{X}_3] = \hat{X}_2 - a \hat{X}_1,\quad
[\hat{X}_2, \hat{X}_3] = - \hat{X}_1 - a \hat{X}_2.
\end{equation}

\subsection*{Symmetry algebra $\hat{\g}_{3,4}$}

\begin{equation}
\label{X34_1}
\hat{X}_1 = D_1 - i e \mu \sin(q^2),
\end{equation}
\begin{equation}
\label{X34_2}
\hat{X}_2 = \sin(q^1) \tan(q^2) D_1 + \cos(q^1) D_2 + i e \mu \sin(q^1) \cos(q^2),
\end{equation}
\begin{equation}
\label{X34_3}
\hat{X}_3 = \cos(q^1) \tan(q^2) D_1 - \sin(q^1) D_2 + i e \mu \cos(q^1) \cos(q^2),
\end{equation}
\begin{equation}
\label{commX34}
[\hat{X}_1, \hat{X}_2] = \hat{X}_3,\quad
[\hat{X}_2, \hat{X}_3] = \hat{X}_1,\quad
[\hat{X}_3, \hat{X}_1] = \hat{X}_2.
\end{equation}

\subsection*{Symmetry algebra $\hat{\g}_{3,5}$}

\begin{equation}
\label{X35_1}
\hat{X}_1 = D_1 - i e \mu \sin(q^2),
\end{equation}
\begin{equation}
\label{X35_2}
\hat{X}_2 = \sinh(q^1) \tan(q^2) D_1 + \cosh(q^1) D_2 + i e \mu \sinh(q^1) \cos(q^2),
\end{equation}
\begin{equation}
\label{X35_3}
\hat{X}_3 = \cosh(q^1) \tan(q^2) D_1 + \sinh(q^1) D_2 + i e \mu \cosh(q^1) \cos(q^2),
\end{equation}
\begin{equation}
\label{commX35}
[\hat{X}_1, \hat{X}_2] = \hat{X}_3,\quad
[\hat{X}_1, \hat{X}_3] = \hat{X}_2,\quad
[\hat{X}_2, \hat{X}_3] = \hat{X}_1.
\end{equation}

\subsection*{Symmetry algebra $\hat{\g}_{4,1}$}

\begin{equation*}
\hat{X}_1 = D_1,\quad
\hat{X}_2 = D_2,\quad
\hat{X}_3 = - q^2 D_1 + q^1 D_2,\quad
\hat{X}_4 = - q^1 D_1 - q^2 D_2 + D_3,
\end{equation*}
\begin{equation*}
[\hat{X}_1, \hat{X}_2] = 0,\quad
[\hat{X}_1, \hat{X}_3] = \hat{X}_2,\quad
[\hat{X}_1, \hat{X}_4] = - \hat{X}_1,\quad
[\hat{X}_2, \hat{X}_3] = - \hat{X}_1,\quad
[\hat{X}_2, \hat{X}_4] = - \hat{X}_2.
\end{equation*}

\section{Noncommutative integrability of the Klein--Gordon equation on $\dS$}
\label{sec:4}

In this study, integrability of the Klein--Gordon equation \eqref{KGeq} is understood in the sense of the following definition~\cite{Kli01, ShaShi95}.

\begin{definition}
\label{IntDef}
The Klein-Gordon equation \eqref{KGeq} is \textit{integrable} if we can reduce the construction of a basis of its solutions to solving an ordinary differential equation.
\end{definition}

Suppose that Eq.~\eqref{KGeq} admits some $(n + 1)$-dimensional first-order symmetry algebra $\hat{\g}$ with the basis
\begin{equation}
\label{basisX}
\hat{X}_0 = i e,\
\hat{X}_1 = X_1^a(x) D_a + i e \chi_1(x),\
\dots,\
\hat{X}_n = X_n^a(x) D_a + i e \chi_n(x).
\end{equation}
As shown in Section~\ref{sec:1}, this algebra is isomorphic to a one-dimensional central extension of the $n$-dimensional Lie algebra $\g = \{ X_A = X_A^a(x) \p_a,\ A = 1, \dots, n \}$ preserving  the 2-form $\F$ of the electromagnetic field.
The symmetry operators \eqref{basisX} satisfy the commutation relations
\begin{equation*}
[\hat{X}_A, \hat{X}_B] = C_{AB}^C \hat{X}_C + \mathbf{F}_{AB} \hat{X}_0,
\end{equation*}
where $C_{AB}^C$ are the structure constants of the Lie algebra $\g$, $\mathbf{F}_{AB}$ are the components of a cocycle of $\g$ with values in the trivial $\g$-module $\mathbb{R}$.
In order to avoid some technical difficulties, we assume that the infinitesimal generators $X_A$ of the algebra $\g$ are algebraically independent, that is, there is no symmetrized polynomial $P$ in $n$ non-commuting variables such that $P(X_1, \dots, X_n) \equiv 0$. 

Let us describe a method for reducing the Klein--Gordon equation \eqref{KGeq} to a differential equation including fewer independent variables than the original one.
For this purpose, we adapt the method of noncommutative integration of linear partial differential equations suggested by Shapovalov and Shirokov (see~\cite{ShaShi95, BagBalGitShi02, Kli01}).

The basic ingredient for the noncommutative integration method is the so-called \textit{$\lambda$-representation} of the Lie symmetry algebra $\hat{\g}$, that is, its operator-irreducible representation realized by an $\ind \hat{\g}$-parametric family of first-order operators acting in the space of partially holomorphic functions of $(\dim \hat{\g} - \ind \hat{\g})/2$ variables.
Here, the non-negative integer $\ind \hat{\g}$, called the \textit{index} of the Lie algebra $\hat{\g}$, is defined as the dimension of regular orbits of the corresponding coadjoint representation~\cite{Pan03}.

In the case of $\hat{\g}$, the Lie algebra is realized by the operators \eqref{basisX}; the $\lambda$-representation of $\hat{\g}$ is defined by the operators
\begin{equation*}
\label{lcomm}
\hat{\ell}_0 = - i e,\quad
\hat{\ell}_A = a_A^\mu(\lambda)\, \frac{\p}{\p \lambda^\mu} + b_A(\lambda, J),\quad
A = 1, \dots, n,
\end{equation*}
which satisfy the commutation relations
\begin{equation*}
[\hat{\ell}_A, \hat{\ell}_B] = C_{AB}^C \hat{\ell}_C + \mathbf{F}_{AB} \hat{\ell}_0,\quad
A,B = 1, \dots, n,
\end{equation*}
where $\lambda = (\lambda^1, \dots, \lambda^s) \in \mathbb{C}^s$, $s = \left ( \dim \hat{\g} - \ind \hat{\g} \right )/2$, $J = (J_1, \dots, J_l) \in \mathbb{R}^l$, $l = \ind \hat{\g} - 1$. The index $\ind \hat{\g}$, in this case, is calculated by the formula
\begin{equation}
\label{IndDef}
\ind \hat{\g} = \dim \hat{\g} - \sup \limits_{f \in \hat{\g}^*} \rank \| C_{AB}^C f_C + \mathbf{F}_{AB} f_0 \|,
\end{equation}
where $\hat{\g}^*$ is the dual space to the Lie algebra $\hat{\g}$.
The operator irreducibility means that all Casimir invariants of the $\lambda$-representation are multiples of the identity operator.
For convenience, we require that the operators $\hat{\ell}_A$ be skew-symmetric with respect to some measure $d \mu(\lambda)$.
We note that there is an efficient computational algorithm for constructing $\lambda$-representations based on the fact that these representations can be obtained as a quantization result of the Lie--Poisson bracket in Darboux coordinates~\cite{Shi00}.

Now, we consider the system of equations
\begin{equation}
\label{Xleqs}
\hat{X}_A(x, \p_x) \varphi_J(x,\lambda) = - \hat{\ell}_A(\lambda, \p_\lambda, J) \varphi_J(x,\lambda),\quad
A = 1, \dots, n,
\end{equation}
where $\varphi_J(x,\lambda)$ is a function of the variables $x = (x^1, \dots, x^m)$ and $\lambda = (\lambda^1, \dots, \lambda^s)$ depending on the real parameters $J = (J_1, \dots, J_l)$.
We note that this system of equations is compatible because the sets of the operators $\{\hat{X}_A\}$ and $\{\hat{\ell}_A\}$ form representations of the same Lie algebra $\hat{\g}$.
Since $\hat{X}_A$ are symmetry operators of the Klein--Gordon equation \eqref{KGeq}, the space of all solutions of the system \eqref{Xleqs} is invariant under the operator $\hat{H}$.
Solving Eq.~\eqref{Xleqs} by the method of characteristics, we have the general solution in the form
\begin{equation}
\label{RPhi}
\varphi_J(x, \lambda) = e^{R_J(x,\lambda)} \Phi_J(v^1(x,\lambda), \dots, v^{\tilde{m}}(x,\lambda)),
\end{equation}
where $R_J(x,\lambda)$ is some function, and $\Phi_J(v) = \Phi_J (v^1, \dots, v^{\tilde{m}})$ is an arbitrary function of $\tilde{m}$ variables $v^1 = v^1(x,\lambda), \dots, v^{\tilde{m}} = v^{\tilde{m}}(x,\lambda)$ that are characteristics of the system~\eqref{Xleqs}.

As we have already noted, the space of the solutions of Eq.~\eqref{Xleqs} is invariant under $\hat{H}$. 
Substituting \eqref{RPhi} in the Klein--Gordon equation \eqref{KGeq} and multiplying the result by the factor $e^{-R_J(x,\lambda)}$, we arrive at the differential equation for the unknown function $\Phi_J(v)$:
\begin{equation}
\label{RedEq}
\hat{\tilde{H}}(v, \p_v, J) \Phi_J(v) = 0,
\end{equation}
where $\hat{\tilde{H}}(v, \p_v, J)$ is a second-order linear differential operator.
The differential equation \eqref{RedEq} is called \textit{the reduced equation}.
It is clear that the reduced equation involves fewer independent variables than the original Klein--Gordon equation~\eqref{KGeq}. 
Thus, we have reduced Eq.~\eqref{KGeq} for $\varphi$ as a function of $(q,u)$ to the ordinary differential equation \eqref{RedEq} for $\Phi_J$ as a function of $v$.

Let us now calculate the number $\tilde{m}$ of independent variables in the reduced equation, which is equal to the number of functionally independent characteristics of the system of equations~\eqref{Xleqs}.
Due to the algebraic independence of the generators $X_A = X_A^a(x) \p_{x^a}$, this number is obviously equal to
\begin{equation*}
\tilde{m} = m + \frac{1}{2} \left ( \dim \hat{\g} - \ind \hat{\g} \right ) - \left ( \dim \hat{\g} - 1 \right ) = m - \frac{1}{2} \left ( \dim \hat{\g} + \ind \hat{\g} \right ) + 1.
\end{equation*}
If $\tilde{m} \leq 1$, then the reduced equation \eqref{RedEq} is an algebraic or ordinary differential equation.
Thus, if the symmetry algebra $\hat{\g}$ satisfies the condition
\begin{equation}
\label{IntCond}
\dim \hat{\g} + \ind \hat{\g} \geq 2 m,
\end{equation}
then the Klein--Gordon equation \eqref{KGeq} is integrable in the sense of Definition \ref{IntDef}.

If the number of independent variables in the reduced equation is equal to $\tilde{m} = 1$, then,  substituting a solution of Eq.~\eqref{RedEq} into \eqref{RPhi}, we obtain a family of solutions of Eq.~\eqref{KGeq} depending on $s = \dim \hat{\g} - m$ parameters $\lambda = (\lambda^1, \dots, \lambda^s)$ and $l = 2 m - 1 - \dim \hat{\g}$ parameters $J = (J_1, \dots, J_l)$.
This family can be chosen as a basis of the solution space of the Klein--Gordon equation.

If $\tilde{m} = 0$, the reduced equation \eqref{RedEq} is purely algebraic and can be considered as the restriction imposed on the parameters $J = (J_1, \dots, J_l)$. 
In this case, the number $s$ of variables $\lambda$ is $s = \dim \hat{\g} - m - 1$, whereas the number of functionally independent parameters $J$ is equal to $l - 1 = 2 m - \dim \hat{\g}$.
As a result, the set of the functions $\varphi_J(x,\lambda) = C \cdot e^{R_J(x,\lambda)}$, where $C$ is a constant, forms a basis of the solution space of the Klein--Gordon equation.

Based on the above theory, let us select all the integrable cases for the Klein--Gordon equation~\eqref{KGeq} on 3D de Sitter space.
In order to do this, we must verify the condition \eqref{IntCond} for each of the symmetry algebras $\hat{\g}_{n,k}$ given in Section \ref{sec:3}.
Since $m = \dim \dS = 3$, this condition takes the form
\begin{equation}
\label{dSIntCond}
\dim \hat{\g} + \ind \hat{\g} \geq 6.
\end{equation}
The results of this verification are summarized in Table \ref{tab:3}, which also contains the dimensions and the indices of algebras $\hat{\g}_{n,m}$ and the numbers $s$, $l$, and $\tilde{m}$ that are the numbers of variables $\lambda$, $J$, and $v$, respectively.
Table \ref{tab:3} shows that the Klein--Gordon equation on $\dS$ is integrable only for electromagnetic fields invariant under three- and four-dimensional subalgebras of the Lie algebra~$\mathfrak{so}(1,3)$.

\begin{table}[!h]
\caption{\small{Verification of the condition \eqref{dSIntCond} for symmetry algebras $\hat{\g}_{n,m}$}.}
\label{tab:3}
\small
\begin{center}
\begin{tabular}{|c|c|c|c|c|c|c|}
\hhline{=======}
Symmetry algebra $\hat{\g}$	&	$\dim \hat{\g}$	& $\ind \hat{\g}$	& $s$ &	$l$	&	$\tilde{m}$	&	Is condition \eqref{dSIntCond} true?	\\
\hhline{=======}
$\hat{\g}_{1,1}$, $\hat{\g}_{1,2}$, $\hat{\g}_{1,3}^a$, $\hat{\g}_{1,4}$	&	
2	&	2	&	0	&	1	&	2	&	No	\\
\hline
$\hat{\g}_{2,1}$, $\hat{\g}_{2,2}$, $\hat{\g}_{2,3}$	&	
3	&	1	&	1	&	0	&	2	&	No	\\
\hline
$\hat{\g}_{3,1}$, $\hat{\g}_{3,2}$, $\hat{\g}_{3,3}^a$, $\hat{\g}_{3,4}$, $\hat{\g}_{3,5}$	&	
4	&	2	&	1	&	1	&	1	&	Yes	\\
\hline
$\hat{\g}_{4,1}$	&	5	&	3	&	1	&	3	&	0	&	Yes\\
\hline
\end{tabular}
\end{center}
\end{table}

Now, we demonstrate how the Klein--Gordon equation on 3D de Sitter space can be reduced to an ordinary differential equation in the integrable cases.
In order to do this, we use the non-commutative integration method described above.
Moreover, we construct solutions of Eq.~\eqref{KGeq} in terms of special functions where possible.
The case of the subalgebra $\g_{4,1}$ corresponds to the absence of an electromagnetic field and is not considered here as trivial.

Since we aim to demonstrate the possibility of integrating the Klein--Gordon equation, here, we restrict ourselves only to the local construction of its exact solutions. 
The global aspect related to investigating the behavior of the solutions on the whole space $\dS$ and the choice of appropriate function space will not be discussed in this study.  

\subsection{Case $\g_{3,1}$}

In the local coordinates $q = (q^1, q^2, q^3) \in \mathbb{R}^3$, determined by \eqref{tr31a}, \eqref{tr31b}, the metric of $\dS$ has the form
\begin{equation}
\label{ds31}
ds^2 = - \exp( 2 q^3 ) \left [ ( d q^1)^2 - (d q^2)^2 \right ] + ( d q^3)^2.
\end{equation}
We note that this coordinate chart covers only the ``half'' of the hyperboloid \eqref{deSit} corresponding the condition $x^3 > 0$.
In the local coordinates $q^a$, the infinitesimal generators of the subalgebra $\g_{3,1}$ have the form
\begin{equation*}
X_1 = \frac{\p}{\p q^1},\quad
X_2 = \frac{\p}{\p q^2},\quad
X_3 = - q^1 \frac{\p}{\p q^1} - q^2 \frac{\p}{\p q^2} + \frac{\p}{\p q^3},
\end{equation*}
and the closed 2-form $\F$ invariant under these vector fields is written as (see Table \ref{tab:2}):
\begin{equation}
\label{F31}
\F = \exp (q^3) (\mu_1 d q^1 + \mu_2 d q^2) \wedge d q^3.
\end{equation}
Here, $\mu_1$ and $\mu_2$ are arbitrary constants.

We choose the electromagnetic potential corresponding to  \eqref{F31} in the form
\begin{equation}
\label{A31}
\A = \exp(q^3) \left ( \mu_1 d q^1 + \mu_2 d q^2 \right ).
\end{equation}
The Klein--Gordon equation \eqref{KGeq} for the metric \eqref{ds31} and the electromagnetic potential \eqref{A31} has the following form
\begin{multline}
\label{KGeq31}
\hat{H} \varphi = \left [ - \exp\left (-2 q^3 \right ) \left ( \frac{\p^2}{ \p (q^1)^2} - \frac{\p^2}{\p (q^2)^2} \right ) + \frac{\p^2}{\p (q^3)^2} + 2\, \frac{\p}{\p q^3} - \right .
\\
\left . - 2 i e \exp(q^3) \left ( \mu_1 q^1 + \mu_2 q^2 \right ) \frac{\p}{\p q^3} - 3 i e \exp(q^3) (\mu_1 q^1 + \mu_2 q^2) - \right . 
\\
\left . - e^2 \exp( 2 q^3 ) \left ( \mu_1 q^1 + \mu_2 q^2 \right )^2 + 6 \zeta + m^2 \right ] \varphi = 0.
\end{multline}

The symmetry operators for Eq.~\eqref{KGeq31} are given by the formulas \eqref{X31a}.
Taking into account the electromagnetic potential \eqref{A31}, these operators are written in the explicit form as
\begin{equation*}
\hat{X}_1 = \frac{\p}{\p q^1} - i e \mu_1 \exp(q^3),\quad
\hat{X}_2 = \frac{\p}{\p q^2} - i e \mu_2 \exp(q^3),\quad
\hat{X}_3 = - q^1 \frac{\p}{\p q^1} - q^2 \frac{\p}{\p q^2} + \frac{\p}{\p q^3}.
\end{equation*}
Together with the trivial operator $\hat{X}_0 = i e$, these operators define the symmetry algebra $\hat{\g}_{3,1}$, and it follows from the commutation relations \eqref{commX31} that this algebra is isomorphic to the direct sum $\g_{3,1} \oplus \mathbb{R}$ (a trivial central extension). 
Using \eqref{IndDef}, it is easy to verify that $\ind \hat{\g}_{3,1} = 2$.

The $\lambda$-representation of the algebra $\hat{\g}_{3,1}$ acts in the space of functions of one real variable $\lambda$ and depend on one real parameter $J$:
\begin{equation*}
\label{lambda31}
\hat{\ell}_0 = - i e,\quad
\hat{\ell}_1 = i J \lambda,\quad
\hat{\ell}_2 = i \lambda,\quad
\hat{\ell}_3 = \lambda\, \frac{\p}{\p \lambda} + \frac{1}{2}.
\end{equation*}
It is easy to see that these operators are skew-symmetric under the usual Lebesgue measure $d\lambda$ on $\mathbb{R}$.
Solving the system of equations \eqref{Xleqs}, we obtain
\begin{equation}
\label{phi31}
\varphi_J \left ( q^1, q^2, q^3, \lambda \right ) = \Phi_J \left ( v \right ) \exp \left [ - i \lambda ( J q^1 + q^2 ) - \frac12\, q^3 + i e \exp(q^3) \left ( \mu_1 q^1 + \mu_2 q^2 \right ) \right ],
\end{equation}
where $v = \lambda \exp(-q^3)$.
Substituting the obtained solution into \eqref{KGeq31}, after some algebra, we obtain an ordinary differential equation for an unknown function $\Phi_J(v)$: 
\begin{equation}
\label{RedEq31}
v^2 \Phi_J''(v) + \left [ (J^2 + 1) v^2 - 2 e ( J \mu_1 + \mu_2) v + m^2 + 6 \zeta + e^2 (\mu_1^2 + \mu_2^2) - \frac{3}{4} \right ] \Phi_J(v) = 0.
\end{equation}
Thus, the family of the functions \eqref{phi31}, parameterized by the parameters $J$ and $\lambda$, forms a basis of the solution space for the Klein-Gordon equation \eqref{KGeq31} if the function $\Phi_J(v)$ is a solution of the reduced equation \eqref{RedEq31}.

If instead of the variable $v$, we introduce the new complex variable $z = 2 i \sqrt{J^2 + 1}\, v$, Eq.~\eqref{RedEq31} is reduced to Whittaker's equation
\begin{equation*}
\tilde{\Phi}_J''(z) + \left [ - \frac{1}{4} + \frac{\alpha}{z} + \frac{\frac{1}{4} - \beta^2}{z^2} \right ] \tilde{\Phi}_J(z) = 0,
\end{equation*}
where $\tilde{\Phi}_J(z) = \Phi_J(2 i \sqrt{J^2 + 1}\, v)$, $\alpha = - i e \left ( J \mu_1 + \mu_2 \right )/\sqrt{J^2 + 1}$, $\beta^2 = 1 - m^2 - 6 \zeta - e \left ( \mu_1^2 + \mu_2^2 \right )$.
This equation has the regular singular point $z = 0$ and the irregular singular point $z = \infty$. Its two linearly independent solutions $M_{\alpha,\beta}(z)$ and $W_{\alpha,\beta}(z)$, called the Whittaker functions, can be expressed in terms of confluent hypergeometric functions~\cite{AbrSteRom88}:
\begin{equation*}
M_{\alpha,\beta}(z) = \exp \left ( - \frac{z}{2} \right ) z^{\frac{1}{2} + \beta}\, M \left ( \frac{1}{2} + \beta - \alpha, 1 + 2 \beta, z \right ), 
\end{equation*}  
\begin{equation*}
W_{\alpha,\beta}(z) = \exp \left ( - \frac{z}{2} \right ) z^{\frac{1}{2} + \beta} U \left ( \frac{1}{2} + \beta - \alpha, 1 + 2 \beta, z \right ).
\end{equation*}

\subsection{Case $\g_{3,2}$}

Using the local coordinates $(q,u) = (q^1, q^2, u^1) \in \mathbb{R}^3$ defined by \eqref{tr32a}, \eqref{tr32b}, for the metric of $\dS$ we have
\begin{equation}
\label{ds32}
ds^2 = - \exp(-2 u^1) \left [ (d q^1)^2 - (d q^2)^2 \right ] + (d u^1)^2.
\end{equation} 
As in the previous case, this coordinate chart covers only the ``half'' of the space $\dS$ corresponding to the positive values of the coordinate $x^3$ in $\mathbb{R}^{1,3}$.
In the coordinates $(q^1, q^2, u^1)$, the basic vector fields $X_A \in \g_{3,2}$ and the corresponding electromagnetic field $\F$  have the form (see Table \ref{tab:2}):
\begin{equation*}
X_1 = \frac{\p}{\p q^1},\quad
X_2 = \frac{\p}{\p q^2},\quad
X_3 = - q^2 \frac{\p}{\p q^1} + q^1 \frac{\p}{\p q^2},
\end{equation*}
\begin{equation*}
\F = \mu\, dq^1 \wedge dq^2.
\end{equation*}
Here, $\mu$ is a constant.

Choosing the electromagnetic field potential in the form
\begin{equation}
\label{A32}
\A = \frac{1}{2} \, \mu \left ( q^1 d q^2 - q^2 d q^1 \right ),
\end{equation}
we obtain that the Klein--Gordon equation for the given electromagnetic field and the metric \eqref{ds32} is
\begin{multline}
\label{KGeq32}
\hat{H}\varphi = \left [ - \exp(2u^1) \left ( \frac{\p^2}{\p (q^1)^2} - \frac{\p^2}{\p (q^2)^2} \right ) + \frac{\p^2}{\p (u^1)^2} - 2\, \frac{\p}{\p u^1} - i e \mu \exp(2 u^1) \left ( q^2 \frac{\p}{\p q^1} - q^1 \frac{\p}{\p q^2} \right ) + \right .
\\
\left . + \frac{1}{4}\, e^2 \mu^2 \exp(2 u^1) \left ( (q^1)^2 + (q^2)^2 \right ) + 6 \zeta + m^2 \right ] \varphi = 0.
\end{multline}
The first-order symmetries of this equation are given by the operators \eqref{X32}, which, taking into account~\eqref{A32}, can be written as
\begin{equation*}
\hat{X}_1 = \frac{\p}{\p q^1} - \frac{1}{2}\, i e \mu q^2,\quad
\hat{X}_2 = \frac{\p}{\p q^2} + \frac{1}{2}\, i e \mu q^1,\quad
\hat{X}_3 = - q^2 \frac{\p}{\p q^1} + q^1 \frac{\p}{\p q^2}.
\end{equation*} 
Together with the trivial operator $\hat{X}_0 = i e$, these operators form the symmetry algebra $\hat{\g}_{3,2}$ isomorphic to a one-dimensional central extension of the algebra $\g_{3,2}$.
From the commutation relations \eqref{commX32}, it follows that this extension is indecomposable for $\mu \neq 0$.

The $\lambda$-representation of the algebra $\hat{\g}_{3,2}$ acts in the space of holomorphic functions on the complex plane $\mathbb{C}$:
\begin{equation}
\label{l32}
\hat{\ell}_0 = - i e,\quad
\hat{\ell}_1 = i\, \frac{\p}{\p \lambda} - \frac12\, i e \mu \lambda,\quad
\hat{\ell}_2 = - \frac{\p}{\p \lambda} - \frac12\,e \mu \lambda,\quad
\hat{\ell}_3 = i \lambda\, \frac{\p}{\p \lambda} - i J.
\end{equation}
Here, $\lambda \in \mathbb{C}$, $J \in \mathbb{R}$.
The operators \eqref{l32} are skew-symmetric under the Gaussian measure $\mu(\lambda) = \exp \left ( - \frac{1}{2}\, e |\lambda|^2 \right )$ on $\mathbb{C}$.

The general solution of the system of equations \eqref{Xleqs}, in this case, has the form
\begin{equation}
\label{phi32}
\varphi_J \left ( q^1,q^2,u^1,\lambda \right ) = \Phi_J(v) \left [ q^1 + i \left ( q^2 + \lambda \right ) \right ]^J \exp \left [ \frac{1}{2}\, e \mu \lambda \left ( i q^1 + q^2 \right ) + \frac{1}{4}\, e \mu \left ( (q^1)^2 + (q^2)^2 \right )  \right ],
\end{equation}
where $v = u^1$.
Substituting this function into Eq.~\eqref{KGeq32}, after some algebraic calculations, we obtain the following ordinary differential equation for the unknown function $\Phi_J(v)$:
\begin{equation}
\label{RedEq32}
\Phi_J''(v) - 2 \Phi_J'(v) + \left [ 6 \zeta + m^2 - e \mu \left ( 2 J + 1 \right ) \exp(2 v) \right ] \Phi_J(v) = 0.
\end{equation}
Two linearly independent solutions of this equation can be expressed in terms of Bessel functions of the first and second kind, respectively:
\begin{equation*}
\Phi_J^{(1)}(v) = \exp(v) J_\alpha \left ( i  \exp(v) \sqrt{e \mu (2 J + 1 )} \right ),\quad
\Phi_J^{(2)}(v) = \exp(v) Y_\alpha \left ( i  \exp(v) \sqrt{e \mu (2 J + 1 )} \right ).
\end{equation*}
Here, $\alpha = \sqrt{1 - m^2 - 6 \zeta}$.

Thus, the set of the functions \eqref{phi32}, where $\Phi_J(u)$ is a solution of the reduced equation \eqref{RedEq32}, can be regarded as a basis of the solution space for Eq.~\eqref{KGeq32}.
This basis is parameterized by the numbers $\lambda$ and $J$.

\subsection{Case $\g_{3,3}^a$}

In this case, the rectifying local coordinates $q = (q^1, q^2, q^3) \in \mathbb{R}^3$ on $\dS$ are given by \eqref{tr33a}, \eqref{tr33b}. 
In these coordinates, the metric $ds^2$, the generators $X_A \in \g_{3,3}^a$, and the electromagnetic field $\F$ invariant under the subalgebra $\g_{3,3^a}$  have the form:
\begin{equation*}
ds^2 = - \exp(2 a q^3) \left [ (d q^1)^2 - (d q^2)^2 \right ] + a^2 (d q^3)^2;
\end{equation*}
\begin{equation*}
X_1 = \frac{\p}{\p q^1},\quad
X_2 = \frac{\p}{\p q^2},\quad
X_3 = - \left ( a q^1 + q^2 \right ) \frac{\p}{\p q^1} + \left ( q^1 - a q^2 \right ) \frac{\p}{\p q^2} + \frac{\p}{\p q^3};
\end{equation*}
\begin{equation*}
\F = \exp ( a q^3 )\left [ \mu_1 \cos(q^3) + \mu_2 \sin(q^3) \right ] d q^1 \wedge d q^3 + \exp ( a q^3 ) \left [ \mu_1 \sin(q^3) - \mu_2 \cos(q^3) \right ] d q^2 \wedge d q^3.
\end{equation*}
Here, $\mu_1$, $\mu_2$, and $a > 0$ are arbitrary constants.

Let us fix the electromagnetic field potential in the form
\begin{equation}
\label{A33}
\A = \mu_1 \exp(a q^3) \left [ q^1 \cos(q^3) + q^2 \sin(q^3) \right ]d q^3 + 
     \mu_2 \exp(a q^3) \left [ q^1 \sin(q^3) - q^2 \cos(q^3) \right ]d q^3.
\end{equation}
Then, the corresponding Klein--Gordon equation can be written as
\begin{multline}
\label{KGeq33}
\hat{H} \varphi = \left \{ - \exp(-2 a q^3) \left ( \frac{\p^2}{\p (q^1)^2} - \frac{\p^2}{\p (q^2)^2} \right ) + \frac{1}{a^2}\, \frac{\p^2}{\p (q^3)^2} + \frac{2}{a}\, \frac{\p}{\p q^3} -  \right . 
\\
\left. - \frac{2 i e}{a^2}\, \exp(a q^3) \left [ \left ( \mu_1 q^1 - \mu_2 q^2 \right ) \cos(q^3) + \left ( \mu_2 q^1 + \mu_1 q^2 \right ) \sin(q^3) \right ] \frac{\p}{\p q^3} -  
\right .
\\
\left . - \frac{i e}{a^2}\, \exp(a q^3) \left [ (3 a q^1 + q^2) \mu_1 + ( q^1 - 3 a q^2) \mu_2 \right ] \cos(q^3) - 
\right .
\\
\left . - \frac{i e}{a^2}\, \exp(a q^3) \left [  (3 a q^2 - q^1) \mu_1 + ( 3 a q^1 + q^2) \mu_2 \right ] \sin(q^3) - \right .
\\
\left . - \frac{e^2}{a^2}\, \exp(2 a q^3) \left [ \left ( q^1 \mu_1 - q^2 \mu_2 \right ) \cos(q^3) - \left ( q^1 \mu_2 + q^2 \mu_1 \right ) \sin(q^3) \right ] + 6 \zeta + m^2 \right \} \varphi = 0.
\end{multline}
The symmetry algebra $\hat{\g}^a_{3,3}$ of this equation is generated by the trivial operator $\hat{X}_0 = i e $ and the operators \eqref{X33_1}--\eqref{X33_3}, which in view of Eq.~\eqref{A33} can be written in the form
\begin{equation*}
\hat{X}_1 = \frac{\p}{\p q^1} - \frac{i e \exp(a q^3)}{1 + a^2} \left [ \left ( a \mu_1 - \mu_2 \right ) \cos(q^3) + \left ( \mu_1 + a \mu_2 \right ) \sin(q^3) \right ],
\end{equation*}
\begin{equation*}
\hat{X}_2 = \frac{\p}{\p q^2} + \frac{i e \exp(a q^3)}{1 + a^2} \left [ \left ( \mu_1 + a \mu_2 \right ) \cos(q^3) - \left ( a \mu_1 - \mu_2 \right ) \sin(q^3) \right ],
\end{equation*}
\begin{equation*}
\hat{X}_3 = - ( a q^1 + q^2 )\frac{\p}{\p q^1} + (q^1 - a q^2)\frac{\p}{\p q^2} + \frac{\p}{\p q^3}.
\end{equation*}
As follows from the commutation relations \eqref{commX33}, $\hat{\g}^a_{3,3}$ is isomorphic to the direct sum of the algebra $\g_{3,3}$ and the one-dimensional center $\langle \hat{X}_0 \rangle \simeq \mathbb{R}$.

The $\lambda$-representation of the Lie algebra $\hat{\g}^a_{3,3}$ acts in the space of functions of one real variable~$\lambda$:
\begin{equation*}
\hat{\ell}_1 = i J \exp(a \lambda) \cos \lambda,\quad
\hat{\ell}_2 = i J \exp(a \lambda) \sin \lambda,\quad
\hat{\ell}_3 = \frac{\p}{\p \lambda}.
\end{equation*}
The general solution of the system of equations \eqref{Xleqs} is given by
\begin{multline*}
\varphi_J(q^1, q^2, q^3, \lambda) = \Phi_J(v) \exp \left [ i e \exp(a q^3) \left ( \frac{a \mu_1 - \mu_2}{1 + a^2}\, q^1 - \frac{\mu_1 +  a \mu_2}{1 + a^2}\, q^2 \right ) \cos(q^3) +
\right . \\ \left . +
 i e \exp(a q^3) \left ( \frac{\mu_1 + a \mu_2}{1 + a^2}\, q^1 + \frac{a \mu_1 - \mu_2}{1 + a^2}\, q^2 \right ) \sin(q^3) - i J \exp(a \lambda) \left ( q^1 \cos(\lambda) + q^2 \sin(\lambda) \right ) \right ],
\end{multline*}
where $v = q^3 - \lambda$.
Substituting it into the Klein-Gordon equation \eqref{KGeq33}, we obtain the reduced equation for the unknown function $\Phi_J(v)$:
\begin{multline}
\label{RedEq33}
\Phi_J''(v) + 2 a \Phi_J'(v) + \left [ - \frac{2 e a^2 J \exp(- a v)}{1 + a^2} \left ( a \mu_1 - \mu_2 \right ) \cos(v) - \frac{2 e a^2 J \exp(- a v)}{1 + a^2} \left ( \mu_1 + a \mu_2 \right ) \sin(v) + 
\right . \\ \left . + 
a^2 J^2 \exp(-2 a v) + a^2 ( m^2 + 6 \zeta ) + \frac{e^2 a^2 (\mu_1^2 + \mu_2^2)}{1 + a^2} \right ] \Phi_J(v) = 0.
\end{multline}
In general, the solutions of this ordinary differential equation are apparently not expressed in terms of   known special functions.
For specific parameter values, \eqref{RedEq33} can be studied numerically or using asymptotic methods.

\subsection{Case $\g_{3,4}$}

In the local coordinates $(q^1, q^2, u^1)$ that are given by \eqref{tr34a}, the metric of $\dS$ is written as
\begin{equation*}
\label{ds34}
ds^2 = - \cosh^2(u^1) \cos^2(q^2) d (q^1)^2 - \cosh^2(u^1) d (q^2)^2 + d (u^1)^2.
\end{equation*}
We note that these coordinates cover de Sitter space almost everywhere.
In particular, $q^1, q^2$ are spherical coordinates on the spacelike surfaces $x^0 = \mathrm{const}$. 

In the coordinates $(q^1, q^2, u^1)$, the infinitesimal generators $X_A \in \g_{3,4}$ and the invariant closed 2-form $\F$ have the form (see Table \ref{tab:2}):
\begin{equation*}
X_1 = \frac{\p}{\p q^1},\quad
X_2 = \sin (q^1) \tan (q^2)\, \frac{\p}{\p q^1} + \cos ( q^1 )\, \frac{\p}{\p q^2},\quad
X_3 = \cos (q^1) \tan (q^2)\, \frac{\p}{\p q^1} - \sin ( q^1 )\, \frac{\p}{\p q^2};
\end{equation*}
\begin{equation}
\label{F34}
\F = \mu \cos ( q^2 ) d q^1 \wedge d q^2,
\end{equation}
where $\mu$ is a constant.
If we choose the electromagnetic field potential in the form
\begin{equation}
\label{A34}
\A = - \mu \sin ( q^2 ) d q^1,
\end{equation}
then the Klein-Gordon equation for this case is written as follows:
\begin{multline}
\label{KGeq34}
\hat{H} \varphi = \left \{ \frac{\p^2}{\p (u^1)^2} + 2 \tanh (u^1) \, \frac{\p}{\p u^1} - \frac{1}{\cosh^2(u^1)} \left ( \frac{1}{\cos^2(q^2)} \, \frac{\p^2}{\p (q^1)^2} + \frac{\p^2}{\p (q^2)^2} - \tan (q^2)\, \frac{\p}{\p q^2} \right ) - 
\right . \\ \left . -
 \frac{2 i e \mu \tan (q^2)}{\cosh^2 (u^1) \cos (q^2)}\, \frac{\p}{\p q^1} + \frac{e^2 \mu^2 \tan^2 (q^2)}{\cosh^2 (u^1)} + 6 \zeta + m^2 \right \} \varphi = 0.
\end{multline}
The first-order symmetry algebra of this equation is generated by the operators \eqref{X34_1}--\eqref{X34_3}, which in view of \eqref{A34} are explicitly given by
\begin{equation*}
\hat{X}_1 = \frac{\p}{\p q^1},\quad
\end{equation*} 
\begin{equation*}
\hat{X}_2 = \sin (q^1) \tan (q^2) \frac{\p}{\p q^1} + \cos (q^1) \frac{\p}{\p q^2} + \frac{i e \mu \sin (q^1)}{\cos (q^2)},
\end{equation*}
\begin{equation*}
\hat{X}_3 = \cos (q^1) \tan (q^2) \frac{\p}{\p q^1} - \sin (q^1) \frac{\p}{\p q^2} + \frac{i e \mu \cos (q^1)}{\cos (q^2)}.
\end{equation*}
Together with the trivial operator $\hat{X}_0 = i e$, these operators form the Lie algebra $\hat{\g}_{3,4}$ isomorphic to the direct sum $\g_{3,4} \oplus \mathbb{R}$ (it can be easily seen from  \eqref{commX34} and, in fact, follows from the semisimplicity of the Lie algebra $\g_{3,4} \simeq \mathfrak{so}(3)$).

We realize the $\lambda$-representation of the Lie algebra $\g_{3,4}$ by the following family of operators
\begin{equation}
\label{l34}
\hat{\ell}_1 = - i \lambda\, \frac{\p}{\p \lambda} + i J,\quad
\hat{\ell}_2 = \frac{i}{2} \left ( 1 - \lambda^2 \right ) \frac{\p}{\p \lambda} + i J \lambda,\quad
\hat{\ell}_3 = - \frac{1}{2} \left ( 1 + \lambda^2 \right ) \frac{\p}{\p \lambda} + J \lambda.
\end{equation}
Here $\lambda \in \mathbb{C}$, $J \in (0, + \infty)$.
The measure under which these operators are skew-symmetric is determined by the equality $d\mu(\lambda) = (1+|\lambda|^2)^{2(J+1)} d \lambda \wedge d\bar{\lambda}$.
Solving the system of equations \eqref{Xleqs}, we find
\begin{multline}
\label{phi34}
\varphi_J(q^1, q^2, u^1, \lambda) = \Phi_J(v) \left [ ( \lambda^2 \exp(i q^1) + \exp(-i q^1)) \cos(q^2) - 2 i \lambda \sin(q^2) \right ]^J \times
\\
\times \left [ \frac{(i \lambda \exp(i q^1) \cos(q^2) + \sin(q^2) + 1)(\sin(q^2)-1)}{(i \lambda \exp(i q^1) \cos(q^2) + \sin(q^2) - 1)\cos(q^2)} \right ]^{e\mu},
\end{multline}
where $v = u^1$.
Substituting this function into the Klein--Gordon equation \eqref{KGeq34}, we obtain the ordinary differential equation for the unknown function $\Phi_J(u)$:
\begin{equation}
\label{RedEq34}
\Phi_J''(v) + 2 \tanh(v) \Phi'_J(v) + \left ( m^2 + 6 \zeta + \frac{J(J+1) - e^2 \mu^2}{\cosh^2(v)} \right ) \Phi_J(v) = 0.
\end{equation}
Two linearly independent solutions of this equation are expressed in terms of associated Legendre functions of the first and second kind:
\begin{equation*}
\Phi_J^{(1)}(v) = \frac{1}{\sqrt{\cosh(v)}} \, P_\nu^\sigma(\tanh(v)),\quad
\Phi_J^{(2)}(v) = \frac{1}{\sqrt{\cosh(v)}} \, Q_\nu^\sigma(\tanh(v)),
\end{equation*}
where the parameters $\nu$ and $\sigma$ are defined as
$$
\nu = \sqrt{\left ( J + \frac12 \right )^2 - e^2 \mu^2} - \frac{1}{2},\quad
\sigma = \sqrt{1 - m^2 - 6 \zeta}.
$$
Thus, the family of functions \eqref{phi34} parameterized by the numbers $J$ and $\lambda$ forms a basis of the solution space of Eq.~\eqref{KGeq34} if the function $\Phi_J(v)$ is a solution to the ordinary differential equation~\eqref{RedEq34}.

\subsection{Case $\g_{3,5}$}

In the local coordinates \eqref{tr35a} and \eqref{tr35b}, the metric of $\dS$ is written as
\begin{equation*}
ds^2 = - \sin^2(u^1) \cos^2(q^2) d (q^1)^2 + \sin^2(u^1) d (q^2)^2 + d (u^1)^2.
\end{equation*}
The basic vector fields $X_A \in \g_{3,5}$ have the form (see Table \ref{tab:2})
\begin{equation*}
X_1 = \frac{\p}{\p q^1},\quad
X_2 = \sinh(q^1) \tan(q^2) \frac{\p}{\p q^1} + \cosh(q^1) \frac{\p}{\p q^2},\quad
X_3 = \cosh(q^1) \tan(q^2) \frac{\p}{\p q^1} + \sinh(q^1) \frac{\p}{\p q^2},
\end{equation*}
and the corresponding closed invariant 2-form $\F$ is given by the expression \eqref{F34}.

Choosing the electromagnetic field potential in the form \eqref{A34}, we obtain the following Klein-Gordon equation:
\begin{multline}
\label{KGeq35}
\hat{H} \varphi = \left [ \frac{\p^2}{\p (u^1)^2} + 2 \cot(u^1) \, \frac{\p}{\p u^1} - \frac{1}{\sin^2(u^1)} \left ( \frac{1}{\cos^2(q^2)}\, \frac{\p^2}{\p (q^1)^2} - \frac{\p^2}{\p (q^2)^2} + \tan(q^2)\, \frac{\p}{\p q^2} \right ) - \right .
\\
\left . - \frac{2 i e \mu \tan(q^2)}{\sin^2(u^1) \cos(q^2)}\, \frac{\p}{\p q^1} + \frac{e^2 \mu^2 \tan^2(q^2)}{\sin^2(u^1)} + 6 \zeta + m^2 \right ] \varphi = 0.
\end{multline}
The symmetry operators \eqref{X35_1}--\eqref{X35_3}, for the selected electromagnetic potential having the form
\begin{equation*}
\hat{X}_1 = \frac{\p}{\p q^1},\quad
\hat{X}_2 = \sinh(q^1) \tan(q^2)\, \frac{\p}{\p q^1} + \cosh(q^1) \frac{\p}{\p q^2} + \frac{i e \mu \sinh(q^1)}{\cos(q^2)},
\end{equation*} 
\begin{equation*}
\quad
\hat{X}_3 = \cosh(q^1) \tan(q^2)\, \frac{\p}{\p q^1} + \sinh(q^1)\, \frac{\p}{\p q^2} + \frac{i e \mu \cosh(q^1)}{\cos(q^2)},
\end{equation*}
together with the trivial symmetry operator $\hat{X}_0 = i e$ form the Lie algebra $\hat{\g}_{3,5}$ isomorphic to the direct sum $\g_{3,5} \oplus \mathbb{R}$. 

It was shown in \cite{Shi00} that the Lie algebra $\mathfrak{so}(1,2)$ has the two different $\lambda$-representations corresponding to discrete and continuous series of representations of the group $SO(1,2)$.
As an example, we choose the representation corresponding to a continuous series (the second case is analyzed similarly and simply leads to a different basis for solutions to the Klein - Gordon equation):
\begin{equation*}
\hat{\ell}_1 = \lambda\, \frac{\p}{\p \lambda} + i J + \frac{1}{2},\quad
\hat{\ell}_2 = \frac{1}{2} \left ( \lambda^2 + 1 \right ) \frac{\p}{\p \lambda} + \left ( i J + \frac{1}{2} \right ) \lambda,\quad
\hat{\ell}_3 = \frac{1}{2} \left ( \lambda^2 - 1 \right ) \frac{\p}{\p \lambda} + \left ( i J + \frac{1}{2} \right ) \lambda.
\end{equation*}
Here, $\lambda \in \mathbb{R}$, $J \in [0, +\infty)$.
These operators are skew-symmetric under the Lebesgue measure $d \lambda$ on $\mathbb{R}$.

It is easy to obtain the general solution of the system of equations \eqref{Xleqs} in this case:
\begin{multline*}
\varphi_J(q^1,q^2,u^1,\lambda) = \Phi_J(v) \left [ 2 \lambda \sin(q^2) + \left ( \exp(q^1) - \lambda^2 \exp(-q^1) \right ) \cos(q^2) \right ]^{-i J - \frac{1}{2}} \times
\\
\times \left [ \frac{\lambda \exp(-q^1) \cos(q^2) + 1 - \sin(q^2)}{\cos(q^2) - \lambda \exp(-q^1) \left ( 1 - \sin(q^2) \right )} \right ]^{i e \mu}.
\end{multline*}
Substituting it into \eqref{KGeq35}, after a series of algebraic transformations, we obtain the ordinary differential equation for the unknown function $\Phi_J(v)$:
\begin{equation*}
\Phi''_J(v) + 2 \cot(v) \Phi'_J(v) + \left [ m^2 + 6 \zeta + \frac{J^2 - e^2 \mu^2 + \frac{1}{4}}{\sin^2(v)} \right ] \Phi_J(v) = 0.
\end{equation*}
The two linearly independent solutions of this equation are expressed in terms of associated Legendre functions of the first and second kind:
\begin{equation*}
\Phi_J^{(1)}(v) = \frac{1}{\sqrt{\sin(v)}} \, P_\nu^\sigma(\cos(v)),\quad
\Phi_J^{(2)}(v) = \frac{1}{\sqrt{\sin(v)}} \, Q_\nu^\sigma(\cos(v)),
\end{equation*}
where the parameters $\nu$ and $\sigma$ are defined as
\begin{equation*}
\nu = \sqrt{1+m^2+6 \zeta} - \frac{1}{2},\quad
\sigma = \sqrt{e^2 \mu^2 - J^2}.
\end{equation*}

\section*{Conclusion}

Employing the defining equations for the first-order symmetry operators of the Klein--Gordon equation in an external electromagnetic field~\cite{Car77, Van07}, we have given an algorithm for constructing these operators and described the structure of the corresponding Lie algebra $\hat{\g}$ in terms of Lie algebra extensions.
In particular, we have shown that $\hat{\g}$ is a one-dimensional central extension of the subalgebra $\g$ of the Killing vector fields that preserve the electromagnetic field tensor.
These results allowed us to suggest a symmetry-based approach to the classification of electromagnetic fields that admit nontrivial first-order symmetry algebras of the Klein--Gordon equation on 3D de Sitter background.
Within the approach, we have obtained the list of all such electromagnetic fields on $\dS$ (see Table~\ref{tab:2}) based on the well-known classification of inequivalent subalgebras of the Lie algebra $\mathfrak{so}(1,3)$. Also, we have explicitly constructed the corresponding first-order symmetry algebras $\hat{\g}$.
We emphasize that all computations were carried out in local coordinates in which the basis vector fields of subalgebras $\g \in \mathfrak{so}(1,3)$ have the simplest form. An original algebraic method for constructing such coordinates is presented in Appendix. 

In the second part of the paper, we have briefly described the method of noncommutative integration 
for linear partial differential equations developed by Shapovalov and Shirokov~\cite{ShaShi95}. 
Unlike the method of separation of variables, the noncommutative integration method more efficiently exploits the first-order symmetry algebra and allows one to integrate the equation without involving second-order symmetry operators. 
Applying the integrability condition \eqref{IntCond} to the constructed first-order symmetry algebras $\hat{\g}$ on $\dS$, we selected the classes of electromagnetic fields admitting the noncommutative integrability of the Klein--Gordon equation (see Table \ref{tab:3}).
All such electromagnetic fields correspond to subalgebras $\g_{3,1}$, $\g_{3,2}$, $\g_{3,3}^a$, $\g_{3,4}$, $\g_{3,5}$, and $\g_{4,1}$ of the Lie algebra $\mathfrak{so}(1,3)$ (see Table \ref{tab:1}).
As the four-dimensional subalgebra $\g_{4,1}$ corresponds to the electromagnetic field with $\F_{ab} = 0$, the Klein-Gordon equation in this electromagnetic field is equivalent to the free Klein--Gordon equation, and we, therefore, did not consider this case.
In contrast, the cases of three-dimensional subalgebras lead to nontrivial electromagnetic fields, and we examined them in detail.
In each case, we reduced the original Klein--Gordon equation \eqref{KGeq} to an auxiliary ordinary differential equation and expressed its general solution in terms of special functions where possible.

\section*{Acknowledgments}

The reported study was funded by RFBR, project number 19-32-90200. Dr. S. V. Danilova is gratefully acknowledged for careful reading of the
manuscript.

\section*{Appendix: An algebraic method for constructing rectifying coordinates}

\setcounter{equation}{0}
\renewcommand\theequation{A.\arabic{equation}}

\renewcommand\thetheorem{A.\arabic{theorem}}

\setcounter{remark}{0}
\renewcommand\theremark{A.\arabic{remark}}

Let $G$ be an $n$-dimensional Lie transformation group  effectively acting on a smooth $m$-dimensional manifold $M$ on the left.
The \textit{orbit} of a point $x_0 \in M$ is a subset $\mathcal{O}_{x_0}$ of $M$ such that $\mathcal{O}_{x_0} = \{ x \in M \colon x = g x_0,\ g \in G \}$. 
Every orbit $\mathcal{O}_{x_0}$ is diffeomorphic to the quotient space $G/G_{x_0}$, where $G_{x_0} = \{ g \in G \colon g x_0 = x_0 \}$ is the stability group of $x_0$, also called the \textit{isotropy subgroup} of~$x_0$.

It is known that the orbits of the group $G$ are immersed submanifolds of~$M$.
A group action is called \textit{semi-regular} if all its orbits have the same dimension. 
The action is called \textit{regular} if, in addition, each point $x \in M$ has a neighborhood whose intersection with each of the orbits is a pathwise connected subset.
It is clear that an action of $G$ on $M$ is semi-regular if and only if the dimension of $G_x$ does not depend on $x \in M$.

\begin{remark}
In general, Lie group actions are not regular or even semi-regular. 
However, if an action of $G$ on $M$ is not semi-regular, then we may restrict our attention to the induced semi-regular action of $G$ on $\tilde{M} = M \setminus S$, where $S$ is the union of all singular orbits of~$G$.
Obviously, $S$ forms a set of measure zero in $M$.
\end{remark}

Manifolds with regular or semi-regular Lie group actions are locally arranged especially simply.
The following fundamental result is a direct consequence of the famous Frobenius' theorem.
 
\begin{theorem}
\label{the:1}
Let the orbits of a semi-regular action of $G$ on $M$ have dimension $r$.
Then, for any point $x_0 \in M$, there exist \textit{rectifying} local coordinates $(q, u) = (q^1, \dots, q^r, u^1, \dots, u^{m - r})$ near $x_0$ such that every orbit intersects the given coordinate chart in the surfaces $u^1 = c_1, \dots, u^{m-r} = c_{m-r}$, where $c_1, \dots, c_{m-r}$ are arbitrary constants.
If, moreover, the action of $G$ on $M$ is regular, then the coordinate chart can be chosen so that each orbit intersects it in at most one such surface.
\end{theorem}

The proof and detailed discussion of this theorem can be found in Olver's book~\cite{Olv95}.
As a consequence of Theorem~\ref{the:1}, the decomposition of $M$ into the orbits of a semi-regular action forms a foliation of~$M$.

In practice, rectifying coordinates $(q,u)$ are usually constructed using the infinitesimal technique.

Let $e_1, \dots, e_n$ be a basis in the Lie algebra $\g$ of the Lie group $G$.
Each basic element $e_A \in \g$ can be associated with an infinitesimal generator $X_A$ of the action $G$ on $M$ according to the formula
\begin{equation*}
\label{ap1:01}
X_A(x) = \frac{d}{dt} \Exp(t e_A) x\, \Big|_{t = 0},\quad
A = 1, \dots, n.
\end{equation*}
Here, $x \in M$, $\Exp: \g \to G$ is the exponential map.
These infinitesimal generators satisfy the commutation relations
\begin{equation}
\label{ap1:02}
[X_A, X_B] = C_{AB}^C X_C,\quad
A,B = 1, \dots, n,
\end{equation}
where the constants $C_{AB}^C$ are the \textit{structure constants} of the Lie algebra~$\g$.
Thus, the correspondence $e_A \to X_A$ extended by linearity to the whole algebra $\g$ is a Lie algebra homomorphism. 
In the case of an effective action, this homomorphism is an isomorphism.
Since the action of the group $G$ on $M$ is supposed to be effective, we will identify the Lie algebra $\g$ with the algebra of infinitesimal generators~$X_A$.

The infinitesimal generators $X_1, \dots, X_n$ at the point $x_0 \in M$ form a basis of the tangent space $T_{x_0} \mathcal{O}_{x_0}$ to the orbit $\mathcal{O}_{x_0}$ at~$x_0$.
If we introduce the notation
\begin{equation}
\label{ap1:03}
r \equiv \dim \mathrm{span}\,(X_1, \dots, X_n)|_{x_0},
\end{equation}
this means that $\dim \mathcal{O}_{x_0} = r$ or, equivalently, $\dim G_{x_0} = n - r$. 
Obviously, if the group $G$ acts on $M$ semi-regularly, then $r$ does not depend on the choice of~$x_0$.
We also note that the isotropy subgroup $G_{x_0}$ is generated by those infinitesimal generators from $\g$ that vanish at~$x_0$; the corresponding subalgebra $\g_{x_0} \subset \g$ is called the \textit{isotropy subalgebra} of the point $x_0$:
\begin{equation*}
\label{ap1:04}
\g_{x_0} = \{ c^A e_A \in \g \colon c^A X_A^a(x_0) = 0,\ c^A \in \mathbb{R} \}.
\end{equation*} 

Now, we describe a general scheme for constructing rectifying coordinates $(q,u)$ near a point $x_0 \in M$.
Let us first make a suitable linear change of coordinates in $M$ and, if necessary, a change of basis in $\g$, so that
\begin{equation}
\label{ap1:05}
X_a(x_0) = \frac{\p}{\p x^a},\quad
a = 1, \dots, r;\quad
X_\alpha(x_0) = 0,\quad
\alpha = r + 1, \dots, n.
\end{equation}
From \eqref{ap1:05}, it follows that the generators $X_{r+1}, \dots, X_n$ form a basis of the isotropy algebra~$\g_{x_0}$.
We consider an $(m-r)$-dimensional submanifold $S_{x_0} \subset M$ formed by the points of the form $s(u) = (x_0^1, \dots$, $x_0^r, x_0^{r+1}+u^1, \dots, x_0^{m}+u^{m-r})$, where $u = (u^1, \dots, u^{m-r}) \in \mathbb{R}^{m-r}$.
By continuity, the orbits of the group $G$ transversally intersect the submanifold $S_{x_0}$ in a neighborhood of $x_0$.
Thus, each point $x$ of the neighborhood can be uniquely written as $x = g s(u)$, where $s(u) \in S_{x_0}$ and $g \in G$.

By virtue of \eqref{ap1:05}, the basis vectors $e_1, \dots, e_r$ span the subspace $\mathscr{M} \subset \g$ that is the complement of the isotropy subalgebra $\g_{x_0}$ in $\g$. 
Since the points $s(u)$ smoothly depend on the parameters $u = (u^1, \dots, u^{m-r})$ and $s(0) = x_0$, the direct sum decomposition $\g = \mathscr{M} \oplus \g_{s(u)}$ is correct for all small values of $u$.
This allows us to write the group element $g \in G$ in the canonical coordinates of the second kind as
$$
g = \prod_{a = 1}^r \Exp(q^a e_a) \cdot \prod_{\alpha = r+1}^{n} \Exp(h^\alpha e_\alpha(u)),\quad 
\text{(no summation over $a$ and $\alpha$)},
$$
where $e_{r+1}(u), \dots, e_{n}(u)$ is a basis of the isotropy subalgebra $\g_{s(u)}$. 
Taking into account that the transformation corresponding to the group element $\Exp(h^\alpha e_\alpha(u))$ leaves the point $s(u)$ fixed, the decomposition $x = gs(u)$ can be rewritten in the form:
\begin{equation}
\label{ap1:06}
x = \Exp(q^1 e_1) \dots \Exp(q^r e_r) s(u).
\end{equation}
The formula \eqref{ap1:06} gives a local diffeomorphism between $(x^1, \dots, x^m)$ and $(q^1, \dots, q^r, u^1, \dots, u^{m-r})$ that determines the desired transformation to rectifying coordinates.

\begin{remark}
Note that in the coordinate chart constructed according to \eqref{ap1:06}, the vector field $X_1$ is automatically ``rectifed'', i.e., $X_1 = \p_{q^1}$.
\end{remark}

In the general case, the explicit calculation of the transformation $x \mapsto \Exp(t e_A) x$ is reduced to finding the integral trajectory of the vector field $X_A$ passing through the point $x$.  
This, in turn, leads to a system of autonomous first-order differential equations. 
However, if the Lie group $G$ acts \textit{linearly}, one can avoid the procedure of integrating differential equations and perform the required calculations, using only linear algebra tools.
Indeed, for a linear action of $G$ on $M$ the components of the corresponding infinitesimal generators can be chosen to be linear and homogeneous functions of $x^a$:
\begin{equation*}
\label{ap1:07}
X_A = - \rho_{Aa}^b x^a \p_{x^b},\quad
A = 1, \dots, n,
\end{equation*}
where $\rho_{Aa}^b$ are constants.
In this case, from the commutation relations \eqref{ap1:02}, it follows that the matrices $\rho_A \equiv \| \rho_{Aa}^b \|$ satisfy the relations
\begin{equation*}
\label{ap1:08}
[\rho_A, \rho_B] = C_{AB}^C \rho_C,\quad
A,B = 1, \dots, n,
\end{equation*}
which mean that the mapping $e_A \mapsto \rho_A$ defines a representation of the Lie algebra $\g$.

The linearity of the generators $X_A$ with respect to coordinates allows us to write the action of the group element $\Exp(t e_A)$ on the point $x \in M$ in the form $e^{-t \rho_A} x$, where $e^Y = \sum_{k=0}^\infty Y^k/k!$ denotes the matrix exponential of a matrix~$Y$.
Thus, the transformation $x = x(q,u)$ to rectifying coordinates can be made by the formula \eqref{ap1:06} with $\Exp(q^a e_a)$ replaced by $e^{-q^a \rho_a}$.
We note that the calculation of matrix exponentials reduces to purely algebraic manipulations and can be carried out using specialized mathematical software (Maple, Wolfram Mathematica, etc.). 

Note that the algebraic construction of rectifying coordinates can also be performed for a non-linear action of the Lie group $G$ if it is induced by a linear semi-regular action of $G$ on some extended manifold.
Indeed, let the Lie group~$G$ act linearly and semi-regularly on a manifold $\tilde{M}$ and let $M \subset \tilde{M}$ be a submanifold of the form
\begin{equation*}
\label{ap1:09}
M = \{ x \in \tilde{M} \colon J_\nu(x) = 0,\quad \nu = 1, \dots, \tilde{m}-m \},
\end{equation*}
where $\{ J_\nu(x) \}$ is the set of functionally independent invariants of the action of $G$ on $\tilde{M}$, $m = \dim M$, $\tilde{m} = \dim \tilde{M}$.
Let $(q, \tilde{u}) = (q^1, \dots, q^r, \tilde{u}^1, \dots, \tilde{u}^{\tilde{m}-r})$ be rectifying local coordinates in a neighborhood of $x_0 \in M$.
Since the functions $J_\nu(x)$ are invariants of the action of $G$, we have $J_\nu(x(q,\tilde{u})) = \kappa_\nu(\tilde{u})$, that is, in the rectifying coordinates, $J_\nu$ depend only on the variables $\tilde{u}^\nu$, $\nu = 1, \dots, \tilde{m}-r$.
Equating functions $\kappa_\nu(\tilde{u})$ to zero and taking into account their functional independence, we find $\tilde{u}^\nu = \tilde{u}^\nu(u^1, \dots, u^{m-r})$, where $\{ u^1, \dots, u^{m-r} \}$ is some collection of new parameters.
The variables $(q^1, \dots, q^r, u^1, \dots, u^{m-r})$ obtained in this way can be taken to be rectifying local coordinates on the invariant submanifold $M$. 

Let us give an example.  
Consider the standard linear action of the group $G = SO(1,2)$ on Minkowski space $\mathbb{R}^{1,3}$ generated by the infinitesimal generators:
\begin{equation}
\label{ap1:11}
X_1 = - x^0 \frac{\p}{\p x^1} - x^1 \frac{\p}{\p x^0},\quad
X_2 = - x^0 \frac{\p}{\p x^2} - x^2 \frac{\p}{\p x^0},\quad
X_3 = - x^2 \frac{\p}{\p x^1} + x^1 \frac{\p}{\p x^2}.
\end{equation}
The commutation relations between these generators are
\begin{equation*}
\label{ap1:12}
[X_1, X_2] = X_3,\quad
[X_2, X_3] = - X_1,\quad
[X_1, X_3] = X_2.
\end{equation*}
The corresponding Lie algebra $\g$ is isomorphic to the semisimple algebra $\mathfrak{so}(1,2)$ and is a subalgebra of the Lie algebra $\mathfrak{so}(1,3)$ (see the subalgebra $\g_{3,5}$ in Table \ref{tab:1}). 
We note that the orbits of $G$ in $\mathbb{R}^{1,3}$ are two-dimensional, except the singular orbits that are points of the line $L = \{ x \in \mathbb{R}^{1,3} \colon x^1 = x^2 = x^0 = 0 \}$. 
Thus, the action is semi-regular (and even regular) on the open set $\tilde{M} = \mathbb{R}^{1,3}\setminus L$, and, in accordance with \eqref{ap1:03}, we have $r = 2$.

The function $J(x) = (x^1)^2 + (x^2)^2 + (x^3)^2 - (x^0)^2$ is an invariant of the action since  $X_A J = 0$ for all $A = 1, 2, 3$ . 
This means that the space $\dS = \{ x \in \mathbb{R}^{1,3} \colon J(x) = 1 \}$ regarded as the surface in $\mathbb{R}^{1,3}$ is invariant under the group $G$.
Almost all orbits of the corresponding induced action are one-dimensional submanifolds in $\dS$, except two singular orbits that are points $P_\pm = (0,0,\pm 1,0)$.
Therefore, we have a semi-regular action of the group $G$ on the open set $M = \dS\setminus \{ P_\pm\}$.

Let us fix the point $x_0 = (1, 0,0,0) \in M$ and construct rectifying coordinates $(q^1,q^2,u)$ in some neighborhood of this point.
As the first step, we choose the new basis $X_1' = X_1,\ X_2' = X_3,\ X_3' = X_2$ in the Lie algebra $\g$ and make the change of coordinates in $\tilde{M}$: $y^1 = - x^0,\ y^2 = x^2,\ y^3 = x^3,\ y^0 = x^1$.
Then, in these coordinates, we have
\begin{equation}
\label{ap1:14}
X'_1 = y^0 \frac{\p}{\p y^1} + y^1 \frac{\p}{\p y^0},\quad
X'_2 = y^0 \frac{\p}{\p y^2} - y^2 \frac{\p}{\p y^0},\quad
X'_3 = y^2 \frac{\p}{\p y^1} + y^1 \frac{\p}{\p y^2}.
\end{equation}
At the point $x_0$ whose new coordinates are $y^1 = y^2 = y^3 = 0$, $y^0 = 1$, the infinitesimal generators \eqref{ap1:14} have the form $X'_1(x_0) = \p_{y^1}$, $X'_2(x_0) = \p_{y^2}$, $X'_3(x_0) = 0$.

The vector fields $X'_A = - \rho_{Ai}^j y^i \p_{y^j}$ being infinitesimal generators of the linear action of $G$ on $\tilde{M}$ determine the matrices  $\rho_A = \| \rho_{Ai}^j \|$ forming a finite-dimensional representation of the Lie algebra $\g$ in the space $\mathbb{R}^{1,3}$:
\begin{equation*}
\label{ap1:15}
\rho_1 = \left ( 
\begin{array}{cccc}
0	&	0	&	0	&	-1	\\
0	&	0	&	0	&	0	\\
0	&	0	&	0	&	0	\\
-1	&	0	&	0	&	0	\\
\end{array}
\right  ),\quad
\rho_2 = \left ( 
\begin{array}{cccc}
0	&	0	&	0	&	0	\\
0	&	0	&	0	&	-1	\\
0	&	0	&	0	&	0	\\
0	&	1	&	0	&	0	\\
\end{array}
\right  ),\quad
\rho_3 = \left ( 
\begin{array}{cccc}
0	&	-1	&	0	&	0	\\
-1	&	0	&	0	&	0	\\
0	&	0	&	0	&	0	\\
0	&	0	&	0	&	0	\\
\end{array}
\right  ).
\end{equation*}
According to the formula \eqref{ap1:06}, the functions $y^i(q,\tilde{u})$ defining rectifying coordinates in a neighborhood of $x_0$ are obtained by computing the expression $y = e^{-q^1 \rho_1} e^{-q^2 \rho_2} s(\tilde{u})$, where $s(\tilde{u}) = (0, 0, \tilde{u}^1, \tilde{u}^2 - 1)$.
Carrying out some calculations and returning to the original coordinates $x^i$, we find
\begin{equation}
\label{ap1:16}
x^1 = \left ( \tilde{u}^2 + 1 \right ) \cosh q^1 \cos q^2,\quad
x^2 = \left ( \tilde{u}^2 + 1 \right ) \sin q^2,\quad
x^3 = \tilde{u}^1,\quad
x^0 = - \left ( \tilde{u}^2 + 1 \right ) \sinh q^1 \cos q^2.
\end{equation}
The formulas \eqref{ap1:16} define the transformation to rectifying coordinates on $\tilde{M}$ in a neighborhood of the point $x_0$.

To construct the rectifying local coordinates $(q^1, q^2, u)$ on de Sitter space $\dS$, we substitute \eqref{ap1:16} into the invariant $J(x)$:
\begin{equation*}
\label{ap1:17}
\kappa(\tilde{u}) \equiv J(x(q,\tilde{u})) = (\tilde{u}^1)^2 + \left ( \tilde{u}^2 + 1 \right )^2.
\end{equation*}
The equation $\kappa(\tilde{u}) = 1$ has the parametric family of solutions $\tilde{u}^1 = \cos u$, $\tilde{u}^2 = \sin u - 1$, where $0 \leq u < 2 \pi$.
Substituting these solutions into \eqref{ap1:16}, we arrive at the equalities
\begin{equation}
\label{ap1:18}
x^1 = \cosh q^1 \cos q^2 \sin u,\quad
x^2 = \sin q^2 \sin u,\quad
x^3 = \cos u,\quad
x^0 = - \sinh q^1 \cos q^2 \sin u,
\end{equation}
which give the desired transformation to rectifying coordinates.
In this chart, the coordinates of the point $x_0$ are $q^1 = q^2 = 0$, $u = \pi/2$, and 
\begin{equation*}
\label{ap1:19}
\mathrm{rank}\, \left \| \frac{\p x^i}{\p q^a}, \frac{\p x^i}{\p u} \right \| \Big|_{x_0} = 3
\end{equation*}
at the point $x_0$. 
It means that the variables $q^1$, $q^2$, and $u$ can indeed be considered as local coordinates on $\dS$ near $x_0$.
The domain of the coordinate chart is defined by the inequalities
\begin{equation*}
\label{ap1:20}
-\infty < q^1 < + \infty,\quad
- \frac{\pi}{2} < q^2 < \frac{\pi}{2},\quad
0 < u < \pi.
\end{equation*}

It can be verified by direct calculation that the infinitesimal generators \eqref{ap1:11} in the coordinates $(q^1, q^2, u)$ take the form
\begin{equation*}
\label{ap1:21}
X_1 = \frac{\p}{\p q^1},\quad
X_2 = \sinh(q^1) \tan(q^2) \frac{\p}{\p q^1} + \cosh(q^1) \frac{\p}{\p q^2},\quad
X_3 = \cosh(q^1) \tan(q^2) \frac{\p}{\p q^1} + \sinh(q^1) \frac{\p}{\p q^2}.
\end{equation*}

Let us now list rectifying local coordinates $(q^1, \dots, q^r, u^1, \dots, u^{3-r})$ on de Sitter space $\dS \subset \mathbb{R}^{1,3}$ for each subalgebra $\g_{n,m}$ in Table \ref{tab:1}.

\subparagraph*{Subalgebra $\g_{1,1}$ ($r = 1$):}

\begin{equation*}
\label{tr11a}
x^1 = \cos(u^2),\quad
x^2 = \cos(u^1) \sin(u^2),\quad
x^3 = \sin(u^1) \sin(u^2) \cosh(q^1),\quad
x^0 = - \sin(u^1) \sin(u^2) \sinh(q^1).
\end{equation*}

\subparagraph*{Subalgebra $\g_{1,2}$ ($r = 1$):}

\begin{equation*}
\label{tr12a}
x^1 = - \cosh(u^1) \cos(u^2) \sin(q^1),\quad
x^2 = \cosh(u^1) \cos(u^2) \cos(q^1),\quad
x^3 = \cosh(u^1) \sin(u^2),
\end{equation*}
\begin{equation*}
\label{tr12b}
x^0 = \sinh(u^1).
\end{equation*}

\subparagraph*{Subalgebra $\g_{1,3}^a$ ($r = 1$):}

\begin{equation*}
\label{tr13a}
x^1 = - \cosh(u^1) \cos(u^2) \sin(q^1),\quad
x^2 = \cosh(u^1) \cos(u^2) \cos(q^1),\quad
\end{equation*}
\begin{equation*}
\label{tr13b}
x^3 = \cosh(u^1) \sin(u^2) \cosh(a q^1) - \sinh(u^1) \sinh(a q^1),
\end{equation*}
\begin{equation*}
x^0 = - \cosh(u^1) \sin(u^2) \sinh(a q^1) + \sinh(u^1) \cosh(a q^1).
\end{equation*}

\subparagraph*{Subalgebra $\g_{1,4}$ ($r = 1$):}

\begin{equation*}
\label{tr14a}
x^1 = q^1 \left ( \cosh(u^1) \sin(u^2) - \sinh(u^1) \right ),\quad
x^2 = \cosh(u^1) \cos(u^2),\quad
\end{equation*}
\begin{equation*}
\label{tr14b}
x^3 = \frac{1}{2}\, (q^1)^2 \left ( \sinh(u^1) - \cosh(u^1) \sin(u^2) \right ) + \cosh(u^1) \sin(u^2),
\end{equation*}
\begin{equation*}
\label{tr14c}
x^0 = \frac{1}{2}\, (q^1)^2 \left ( \sinh(u^1) - \cosh(u^1) \sin(u^2) \right ) + \sinh(u^1).
\end{equation*}

\subparagraph*{Subalgebra $\g_{2,1}$ ($r = 2$):}

\begin{equation*}
\label{tr21a}
x^1 = q^1 \exp(-u^1),\quad
x^2 = q^2 \exp(-u^1),\
\end{equation*}
\begin{equation*}
\label{tr21b}
x^3 = \cosh(u) - \frac{1}{2}\, \exp(-u^1) \left ( (q^1)^2 + (q^2)^2 \right ),\quad
x^0 = \sinh(u) - \frac{1}{2}\, \exp(-u^1) \left ( (q^1)^2 + (q^2)^2 \right ).
\end{equation*}

\subparagraph*{Subalgebra $\g_{2,2}$ ($r = 2$):}

\begin{equation*}
\label{tr22a}
x^1 = \cos(u^1) \cos(q^1),\quad
x^2 = \cos(u^1) \sin(q^1),\quad
x^3 = \sin(u^1) \cosh(q^2),\quad
x^0 = - \sin(u^1) \sinh(q^2).
\end{equation*}

\subparagraph*{Subalgebra $\g_{2,3}$ ($r = 2$):}

\begin{equation*}
\label{tr23a}
x^1 = q^1 \exp(q^2) \cos(u^1),\quad
x^2 = \sin(u^1),
\end{equation*}
\begin{equation*}
\label{tr23b}
x^3 =   \cos(u^1) \left ( \cosh(q^2) - \frac{1}{2}\, (q^1)^2 \exp(q^2) \right ),\quad
x^0 = - \cos(u^1) \left ( \sinh(q^2) + \frac{1}{2}\, (q^1)^2 \exp(q^2) \right ).
\end{equation*}

\subparagraph*{Subalgebra $\g_{3,1}$ ($r = 3$):}

\begin{equation}
\label{tr31a}
x^1 = q^1 \exp(q^3),\
x^2 = q^2 \exp(q^3),
\end{equation}
\begin{equation}
\label{tr31b}
x^3 =   \cosh(q^3) - \frac{1}{2}\, \exp(q^3) \left ( (q^1)^2 + (q^2)^2 \right ),\quad
x^0 = - \sinh(q^3) - \frac{1}{2}\, \exp(q^3) \left ( (q^1)^2 + (q^2)^2 \right ).
\end{equation}

\subparagraph*{Subalgebra $\g_{3,2}$ ($r = 2$):}

\begin{equation}
\label{tr32a}
x^1 = q^1 \exp(-u^1),\quad
x^2 = q^2 \exp(-u^1),
\end{equation}
\begin{equation}
\label{tr32b}
x^3 = \cosh(u) - \frac{1}{2}\, \exp(-u) \left ( (q^1)^2 + (q^2)^2 \right ),\quad
x^0 = \sinh(u) - \frac{1}{2}\, \exp(-u) \left ( (q^1)^2 + (q^2)^2 \right ).
\end{equation}

\subparagraph*{Subalgebra $\g_{3,3}^a$ ($r = 3$):}

\begin{equation}
\label{tr33a}
x^1 = q^1 \exp(a q^3),\quad
x^2 = q^2 \exp(a q^3),
\end{equation}
\begin{equation}
\label{tr33b}
x^3 =   \cosh(a q^3) - \frac{1}{2}\, \exp(a q^3) \left ( (q^1)^2 + (q^2)^2 \right ),\
x^0 = - \sinh(a q^3) - \frac{1}{2}\, \exp(a q^3) \left ( (q^1)^2 + (q^2)^2 \right ).
\end{equation}

\subparagraph*{Subalgebra $\g_{3,4}$ ($r = 2$):}

\begin{equation}
\label{tr34a}
x^1 = - \cosh(u^1) \sin(q^1) \cos(q^2),\quad
x^2 = \cosh(u^1) \cos(q^1) \cos(q^2),\quad
x^3 = \cosh(u^1) \sin(q^2),
\end{equation}
\begin{equation}
\label{tr34b}
x^0 = \sinh(u^1).
\end{equation}

\subparagraph*{Subalgebra $\g_{3,5}$ ($r = 2$):}

\begin{equation}
\label{tr35a}
x^1 = \sin(u^1) \cosh(q^1) \cos(q^2),\quad
x^2 = \sin(u^1) \sin(q^2),\quad
x^3 = \cos(u^1),
\end{equation}
\begin{equation}
\label{tr35b}
x^0 = - \sin(u^1) \sinh(q^1) \cos(q^2).
\end{equation}

\subparagraph*{Subalgebra $\g_{4,1}$ ($r = 3$):}

\begin{equation*}
\label{tr41a}
x^1 = q^1 \exp(q^3),\quad
x^2 = q^2 \exp(q^3),\quad
\end{equation*}
\begin{equation*}
\label{tr41b}
x^3 =   \cosh(q^3) - \frac{1}{2}\, \exp(q^3) \left ( (q^1)^2 + (q^2)^2 \right ),\
x^0 = - \sinh(q^3) - \frac{1}{2}\, \exp(q^3) \left ( (q^1)^2 + (q^2)^2 \right ).
\end{equation*}

\bibliographystyle{iopart-num}
\bibliography{biblio}

\providecommand{\newblock}{}
\begin{thebibliography}{10}
\expandafter\ifx\csname url\endcsname\relax
  \def\url#1{{\tt #1}}\fi
\expandafter\ifx\csname urlprefix\endcsname\relax\def\urlprefix{URL }\fi
\providecommand{\eprint}[2][]{\url{#2}}

\bibitem{Fur51}
Furry W 1951 {\em Physical Review\/} {\bf 81} 115

\bibitem{BirDev84}
Birrell N and Davies P 1984 {\em Quantum fields in curved space\/} 7 (Cambridge
  university press)

\bibitem{GriMamMos94}
Grib A~A, Mostepanenko V and Mamayev S 1994 {\em Vacuum quantum effects in
  strong fields\/} (Fridmann Lab.)

\bibitem{ShaShi95}
Shapovalov A~V and Shirokov I 1995 {\em Theoretical and Mathematical Physics\/}
  {\bf 104} 921--934

\bibitem{Mil77}
Miller~Jr W 1977

\bibitem{Woo75}
Woodhouse N 1975 {\em Communications in Mathematical Physics\/} {\bf 44} 9--38

\bibitem{Sha79}
Shapovalov V 1979 {\em Siberian Mathematical Journal\/} {\bf 20} 790--800

\bibitem{Ben16}
Benenti S 2016 {\em Symmetry, Integrability and Geometry: Methods and
  Applications\/} {\bf 12} 013

\bibitem{BagBalGitShi02}
Bagrov V~G, Baldiotti M~C, Gitman D~M and Shirokov I~V 2002 {\em Journal of
  Mathematical Physics\/} {\bf 43} 2284--2305

\bibitem{Kli01}
Klishevich V~V 2001 {\em Classical and Quantum Gravity\/} {\bf 18} 3735

\bibitem{Mag12}
Magazev A~A 2012 {\em Theoretical and Mathematical Physics\/} {\bf 173}
  1654--1667

\bibitem{Ort04}
L{\'{o}}pez-Ortega A 2004 {\em General Relativity and Gravitation\/} {\bf 36}
  1299--1319

\bibitem{All85}
Allen B 1985 {\em Physical Review D\/} {\bf 32} 3136--3149

\bibitem{Yag09}
Yagdjian K and Galstian A 2009 {\em Communications in mathematical physics\/}
  {\bf 285} 293--344

\bibitem{Otc85}
Otchik V~S 1985 {\em Classical and Quantum Gravity\/} {\bf 2} 539--543

\bibitem{Pol89}
Polarski D 1989 {\em Classical and Quantum Gravity\/}  893--900

\bibitem{Gar94}
Garriga J 1994 {\em Physical Review D\/} {\bf 49} 6343--6346

\bibitem{Vil95}
Villalba V~M 1995 {\em Physical Review D\/} {\bf 52} 3742--3745

\bibitem{Mor09}
Moradi S 2009 {\em Modern Physics Letters A\/} {\bf 24} 1129--1136

\bibitem{BavKimSta18}
Bavarsad E, Kim S~P, Stahl C and Xue S~S 2018 {\em Physical Review D\/} {\bf
  97}

\bibitem{Car77}
Carter B 1977 {\em Physical Review D\/} {\bf 16} 3395

\bibitem{Van07}
Van~Holten J 2007 {\em Physical Review D\/} {\bf 75} 025027

\bibitem{de1998lie}
de~Azc{\'a}rraga J~A and Izquierdo J~M 1998 {\em Lie groups, {L}ie algebras,
  cohomology and some applications in physics\/} (Cambridge University Press)

\bibitem{de1998introduction}
de~Azc{\'a}rraga J~A, Izquierdo J~M and Bueno J 1998 {\em arXiv preprint
  physics/9803046\/}

\bibitem{Jac79}
Jacobson N 1979 {\em Lie algebras\/} 10 (Courier Corporation)

\bibitem{FriWin64}
Fris I and Winternitz P 1964 Invariant expansions of relativistic amplitudes
  and subgroups of the proper {L}orentz group Tech. rep. Joint Inst. of Nuclear
  Research, Dubna, USSR Lab. of Theoretical Physics

\bibitem{BarBarFus91}
Barannik A, Barannik L and Fushchich V 1991 {\em Subgroup analysis of
  {G}alilean and {P}oincare groups and reduction of nonlinear equations.\/}
  (Naukova Dumka, Kiev)

\bibitem{Hal04}
Hall G~S 2004 {\em Symmetries and curvature structure in general relativity\/}
  (World Scientific)

\bibitem{Olv00}
Olver P~J 2000 {\em Applications of {L}ie groups to differential equations\/}
  vol 107 (Springer Science \& Business Media)

\bibitem{Pan03}
Panyushev D~I 2003 The index of a {L}ie algebra, the centralizer of a nilpotent
  element, and the normalizer of the centralizer {\em Mathematical Proceedings
  of the Cambridge Philosophical Society\/} vol 134 (Cambridge University
  Press) pp 41--59

\bibitem{Shi00}
Shirokov I~V 2000 {\em Theoretical and Mathematical Physics\/} {\bf 123}
  754--767

\bibitem{AbrSteRom88}
Abramowitz M, Stegun I~A and Romer R~H 1988 Handbook of mathematical functions
  with formulas, graphs, and mathematical tables

\bibitem{Olv95}
Olver P~J 1995 {\em Equivalence, invariants and symmetry\/} (Cambridge
  University Press)

\end{thebibliography}

\end{document}